\pdfoutput=1
\documentclass[epjST]{svjour}
\usepackage[english]{babel}
\usepackage[utf8]{inputenc}
\usepackage{graphicx, epsfig, color}
\usepackage{amsfonts,amssymb,amsmath,latexsym}
\usepackage{subfigure}
\usepackage{cite}
\begin{document}
\title{Complex networks in climate dynamics}
\subtitle{Comparing linear and nonlinear network construction methods}
\author{Jonathan F. Donges\inst{1,2}\fnmsep\thanks{\email{donges@pik-potsdam.de}} \and Yong Zou\inst{1} \and Norbert Marwan\inst{1} \and J\"urgen Kurths\inst{1,3}}
\institute{Potsdam Institute for Climate Impact Research, P.O. Box 60 12 03, 14412 Potsdam, Germany \and Institute of Physics, University of Potsdam, Karl-Liebknecht-Str. 24/25, 14476 Potsdam-Golm, Germany \and Department of Physics, Humboldt University, Newtonstr. 15, 12489 Berlin, Germany}
\abstract{
Complex network theory provides a powerful framework to statistically investigate the topology of local and non-local statistical interrelationships, i.e. teleconnections, in the climate system. Climate networks constructed from the same global climatological data set using the linear Pearson correlation coefficient or the nonlinear mutual information as a measure of dynamical similarity between regions, are compared systematically on local, mesoscopic and global topological scales. A high degree of similarity is observed on the local and mesoscopic topological scales for surface air temperature fields taken from AOGCM and reanalysis data sets. We find larger differences on the global scale, particularly in the betweenness centrality field. The global scale view on climate networks obtained using mutual information offers promising new perspectives for detecting network structures based on nonlinear physical processes in the climate system.
} 
\maketitle
%
%


\section{Introduction \label{Intro}}

During the last decade, the development and application of complex network theory generated a wealth of novel insights into the nature of complex systems in various areas of science, e.g. the internet and world wide web in computer science, food webs, gene expression and neural networks in biology and citation networks in social science \cite{Watts1998,Newman2003,Albert2002}. The intricate interplay between the structure and dynamics of real world networks has received considerable attention \cite{Boccaletti2006}. Particularly, synchronization arising by the transfer of dynamical information in complex network topologies has been studied intensively \cite{Arenas200893}. The application of complex network theory to climate science is a very young field, where only few studies have been reported recently \cite{tsonis2004acn,tsonis2006,tsonis2008jclim,tsonis2008tap,yamasaki2008,gozol2008,Donner2008nts,Donges2008}. The vertices of a climate network are identified with the spatial grid points of an underlying global climate data set. Edges are added between pairs of vertices depending on the degree of statistical interdependence between the corresponding pairs of anomaly time series taken from the climate data set.

When studying networks in the climate system, one has to assume that its dynamics can be approximated reasonably well by a grid of low dimensional nonlinear dynamical systems interacting only with their spatial neighbors according to the locality principle of classical physics. Note that this assumption is made implicitly, when the fundamental partial differential equations of fluid mechanics are discretized and integrated in large scale climate simulations by the coupled atmosphere-ocean general circulation models (AOGCMs) used in climate science. Due to the continuity of the underlying physical fields, such as temperature or pressure, neighboring grid points are dynamically correlated; these trivial local correlations usually decay quickly within a typical length scale. Additionally, richly structured long range correlations appear, that were named teleconnections by the climatological community and have been studied extensively since the end of the nineteenth century \cite{wallace1981tgh}.

The climate network approach enables novel insights into the topology and dynamics of the climate system over many spatial scales disclosed by local network measures, e.g. the number of first neighbors of a vertex $v$ (the degree centrality $k_v$), mesoscopic measures such as the clustering coefficient and global measures, e.g. the average path length. The local degree centrality and related measures have been used to identify super-nodes (regions of high degree centrality) and to associate them with teleconnection patterns in the atmosphere, most notably the North Atlantic Oscillation (NAO) \cite{tsonis2004acn,tsonis2006,tsonis2008jclim}. On the global scale, climate networks were found to possess 'small-world' properties due to long range connections (edges linking geographically very distant vertices), that stabilize the climate system and enhance the information transfer within it \cite{tsonis2004acn,tsonis2006,tsonis2008jclim}. We stress, that the transfer of information in any complex physical system, e.g. the climate system studied here, will be carried by a flow of matter and energy. By studying the prevalence of long range connections in El Ni\~no and La Ni\~na climate networks \cite{tsonis2008tap} and the time dependence of the number of stable edges \cite{yamasaki2008,gozol2008}, it has been shown very recently, that the El Ni\~no-Southern Oscillation (ENSO) has a strong impact on the stability of the climate system.

In all works mentioned above, researchers have used the linear cross-correlation function of pairs of anomaly time series to quantify the degree of statistical interdependence between different spatial regions. But the highly nonlinear processes at work in the climate system call for the application of nonlinear methods to obtain more reliable results. In a recent work on structures in the betweenness centrality field of climate networks \cite{Donges2008}, we have introduced mutual information \cite{kantz2004nts} as a measure of statistical interdependence to climate network construction. The mutual information allows to capture nonlinear relationships between time series. We found that, while many properties of climate networks generated using the Pearson correlation and the mutual information at zero lag are qualitatively and quantitatively similar, the betweenness centrality field shows much greater deviations between the two construction methods. To check the possibility, that these pronounced differences are a signature of nonlinear processes in the climate system, and to bridge the gap between our nonlinear network construction method and the techniques previously used, we present a systematic statistical similarity study of the resulting climate networks. We show, that over a wide range of relevant edge densities (the fraction of the maximum number of possible edges present in the network), a high degree of similarity is maintained on local and mesoscopic topological scales. Furthermore, we address some of the more pronounced differences on the global topological scale, that are uncovered by betweenness centrality, and their possible relation to nonlinear processes in the climate system.

The organization of the paper is the following: We first describe the data and the filtering and normalization procedures applied to it (Sect. \ref{Data}). After introducing the required elements of complex network theory (Sect. \ref{Network_measures}), we proceed to develop in detail the method of climate network construction (Sect. \ref{Construction}). In Sect. \ref{Results}, we present the systematic comparison of the measures obtained from Pearson correlation and mutual information climate networks, respectively. Furthermore we provide a climatological interpretation. Some conclusions are drawn in Sect. \ref{Conclusions}.


\section{Data \label{Data}}

\subsection{Description \label{Data_description}}

We utilize the monthly averaged global surface air temperature (SAT) field for climate network construction to maintain consistency with earlier works that analyzed the same field \cite{tsonis2004acn,tsonis2006,tsonis2008jclim,tsonis2008tap,yamasaki2008,gozol2008,Donges2008}. The SAT field allows to directly capture the complex dynamics on the interface between ocean and atmosphere due to heat exchange and other local processes. SAT therefore enables us to study atmospheric as well as oceanic dynamics within a common framework. We use reanalysis data provided by the National Center for Environmental Prediction/National Center for Atmospheric Research (NCEP/NCAR) \cite{kistler2001nny} and model output from the World Climate Research Programme's (WCRP's) Coupled Model Intercomparison Project phase 3 (CMIP3) multi-model data set \cite{meehl2007wcm}. For optimal comparability with the reanalysis data, we choose a 20th century reference run (20c3m, as defined in the IPCC AR4) by the Hadley Centre HadCM3 model. A data set consists of a regular spatiotemporal grid with time series $x_i(t)$ associated to every spatial grid point $i$ at latitude $\lambda_i$ and longitude $\phi_i$. Start and end dates, length of time series $\mathcal{T}$, latitudinal resolution $\Delta\lambda$, longitudinal resolution $\Delta\phi$ and the number of vertices of the corresponding global climate network $N$ are given in Table \ref{GlobalDatasetsTable}. Note that we remove the polar grid points at $\lambda \in \{-90^\circ,90^\circ\}$ from the data sets, since the poles are represented by rows of grid points with identical dynamics.

\begin{table}
\caption{\label{GlobalDatasetsTable}Properties of global surface air temperature data sets.}
\center
\begin{tabular}{l|c|r}
 & NCEP/NCAR reanalysis & HadCM3\\
 \hline
 \hline
Temporal coverage & Jan 1948 - Dec 2007 & Jan 1860 - Dec 1999\\
$\mathcal{T}$ [months] & 720 & 1680\\
$\Delta\lambda$ [$^\circ$] & 2.5 & 2.5\\
$\Delta\phi$ [$^\circ$] & 2.5 & 3.75\\
$N$ & 10224 & 6816\\
\end{tabular}
\end{table}

\subsection{Filtering and normalization \label{Data_filtering}}

To minimize the bias introduced by the external solar forcing common to all time series in the data set, we calculate anomaly values, i.e. remove the mean annual cycle by phase averaging. Relabeling the time series by month $m \in \left\{1,\dots,12\right\}$ and year $y$ mapping $x_i(t) \rightarrow x_i(y,m)$ one obtains anomaly time series $a_i(y,m) = x_i(y,m) - \left< x_i(y,m)\right>_{y}$, that are consequently subjected to the inverse mapping $a_i(y,m) \rightarrow a_i(t)$. Here and in the following $\left<f(x)\right>_x$ denotes the expectation value of observable $f$ taken with respect to the variable $x$. Note that the anomaly time series already have zero mean. We furthermore normalize the anomaly time series to unit variance. Up to this point, we follow the method used previously by \cite{tsonis2008tap,yamasaki2008}. It is known, that the annual cycle induces higher order effects such as seasonal variability of anomaly time series variance. We find that using only data from a particular season to avoid biases due to this effect does not alter our results substantially, so that we choose to use the whole data set for a more accurate evaluation of statistical interdependence. 


\section{Elements of complex network theory \label{Network_measures}}

Formally, a network or graph is defined as an ordered pair $G: = (V,E)$ containing a set $V=\{1,...,N\}$ of vertices together with a set $E$ of edges $\{i,j\}$, which are 2-element subsets of $V$. In this work we consider undirected and unweighted simple graphs, where only one edge can exist between a pair of vertices and self-loops of the type $\{i,i\}$ are not allowed. This type of graph can be represented by the symmetric adjacency matrix

\begin{equation}
A_{ij} = \begin{cases}
    0 & \{i,j\} \notin E\\
    1 & \{i,j\} \in E.\\
\end{cases}
\end{equation}

\noindent The edge density of a network is given by $\rho = |E| / {N \choose 2} = \left<k_v\right>_v / N$, $|E|$ being the number of edges in the graph and $\left<k_v\right>_v$ the mean vertex degree. The network measures defined below were selected for this study, because they allow us to compare different aspects of climate network topology on local, mesoscopic and global scales (Table \ref{topo_scales_table}) and are well established in the literature \cite{Newman2003,Albert2002,Boccaletti2006,freeman1979csn}. Degree centrality, the related area weighted connectivity and the Hamming distance use only local information on the direct neighborhood of a vertex $v$. In contrast, the closeness and betweenness centralities as well as the average path length include global topological information by relying on shortest paths between pairs of vertices in the network. This is why we refer to the latter three as global measures. On the mesoscopic scale, the local and average clustering coefficient depend only on information about neighbors and next neighbors of vertices. The concept of topological scales is elaborated in greater detail in \cite{GorkaDiss2008}. We refer to measures assigning a real number $g_v \in \mathbb{R}$ to each vertex $v \in V$ via a mapping $V \rightarrow \mathbb{R}: v \mapsto g_v$ as fields. Scalar measures produce a single real number for the whole graph.

Note that for the data sets analyzed here (Sect. \ref{Data}), vertices are not distributed homogeneously on the earth's surface. The density of vertices increases from the equator towards the poles. This induces an inherent bias in the network measures studied, which prompts to use area weighted generalizations of the standard complex network measures, e.g. area weighted connectivity is the generalization of degree centrality. We have performed extensive studies of climate networks constructed from data interpolated to different grids and resolutions and find, that our results (Sect. \ref{Results}) are not altered significantly by the vertex density bias \cite{Sexton2009}. This holds particularly for the highly interesting path based measures on the global topological scale.

\begin{table}
\caption{\label{topo_scales_table}Classification of network measures into topological scales, fields and scalar measures.}
\center
\begin{tabular}{l|c|c|r}
 & local & mesoscopic & global\\
 \hline
 \hline
field & $k_v$, $AWC_v$ & $\mathcal{C}_v$ & $CC_v$, $BC_v$ \\
scalar & $H(A,B)$ & $\mathcal{C}$ &  $\mathcal{L}$ \\
\end{tabular}
\end{table}

\subsection{Local measures \label{Local_measures}}

\subsubsection{Degree centrality}

The degree or degree centrality \cite{freeman1979csn} $k_v$ gives the number of first neighbors of a vertex $v$ and can be calculated from the network adjacency matrix $A_{ij}$ using 

\begin{equation}
k_v = \sum_{i=1}^{N} A_{vi}.
\end{equation}

\noindent Vertices with exceptionally high degree centrality are usually referred to as \emph{hubs} or \emph{super-nodes}. We extend the use of this term to regions of spatially adjacent vertices with high degree centrality.

\subsubsection{Area weighted connectivity}

The area weighted connectivity

\begin{equation}
AWC_v = \frac{ \sum_{i=1}^{N} A_{vi} \cos( \lambda_i ) }{ \sum_{i=1}^{N} \cos( \lambda_i ) },
\end{equation}

\noindent is closely related to the degree centrality $k_v$ of $v$. It corrects for the fact that in geographical networks defined on a grid, vertices correspond to regions of different area on the earth's surface. For the angularly equidistant grids considered in this work, the corresponding area of vertex $v$ is proportional to the cosine of latitude $\lambda_v$ (see Sect. \ref{Data_description}). $AWC_v$ can be interpreted as the fraction of the earth's surface area a vertex is connected to \cite{tsonis2006}. $AWC$ is thus normalized to $0 \leq AWC_v \leq 1$.

\subsubsection{Hamming distance}

The Hamming distance $H(A,B)$ of two labeled simple graphs with adjacency matrices $A_{ij}$ and $B_{ij}$ measures the fraction of edges that have to be changed to transform one graph into the other \cite{hamming26wed}. Both graphs must contain the same number of vertices $N$. Specifically, $H(A,B)$ is given by

\begin{equation}
H(A,B) = \left<XOR(A_{ij}, B_{ij})\right>_{ij},
\end{equation}

\noindent where 

\begin{equation}
XOR(A_{ij}, B_{ij}) = \begin{cases}
    1 & A_{ij} \neq B_{ij}\\
    0 & else.\\
\end{cases}
\end{equation}

\noindent Hamming distance is bounded by $0\leq H(A,B) \leq 1$ and measures the global probability of non-equal entries in the two adjacency matrices. In our application we calculate the Hamming distance of two graphs with approximately equal edge density $\rho$. 

To evaluate the significance of this measurement, we compare it with the expected Hamming distance $H^R(\rho)$ of two independent Erd\H{o}s-R\'enyi random graphs of edge density $\rho$ \cite{erdHos1959rg}. The probability that the entries $A_{ij}$ and $B_{ij}$ differ between the two random graph adjacency matrices is given by $\textrm{p} \left(A_{ij} \neq B_{ij}\right) = \textrm{p} \left(A_{ij}=1 \right) \textrm{p}\left(B_{ij}=0\right) + \textrm{p} \left(A_{ij}=0 \right) \textrm{p} \left(B_{ij}=1\right) = \rho(1-\rho) + (1-\rho)\rho = 2\rho(1-\rho)$. Since all entries are independent, taking the expectation value reveals the expression $H^R(\rho) = \left<\textrm{p}\left(A_{ij} \neq B_{ij}\right)\right>_{ij} = 2\rho(1-\rho)$. The expected Hamming distance $H^R(\rho)$ gives a reference point against which to judge the similarity of two graphs. We will make use of it in Sect. \ref{Global_comparison} to compare the performance of two network measures in climate network construction.

\subsection{Mesoscopic measures \label{Mesoscopic_measures}}

\subsubsection{Local clustering coefficient}

We refer to $\mathcal{C}_v$ as the local topological clustering coefficient or Watts-Strogatz clustering coefficient \cite{Watts1998} of a vertex $v$. It gives the probability, that two randomly chosen first neighbors of $v$ are also neighbors. With $\Gamma_v$ being the set of first neighbors of $v$ and $e(\Gamma_v)$ the number of edges connecting the vertices within the neighborhood $\Gamma_v$, the clustering coefficient can be written as

\begin{equation}
\mathcal{C}_v = \frac{e(\Gamma_v)}{{k_v \choose 2}},
\end{equation}

\noindent where the binomial coefficient ${k_v \choose 2} = \frac{1}{2} k_v (k_v - 1)$ gives the maximum number of edges in $\Gamma_v$. The local clustering coefficient is normalized to $0\leq \mathcal{C}_v \leq 1$.

\subsubsection{Global clustering coefficient}

We speak of the (global) clustering coefficient $\mathcal{C}$ as the mean Watts-Strogatz clustering coefficient

\begin{equation}
\mathcal{C} = \left<\mathcal{C}_v\right>_v.
\end{equation}

\subsection{Global measures\label{Global_measures}}

\subsubsection{Closeness centrality}

Closeness centrality $CC_v$ measures the inverse average topological distance of vertex $v$ to all others in the network \cite{freeman1979csn},

\begin{equation}
CC_v = \frac{N-1}{\sum_{i=1}^N d_{vi}},
\end{equation}

\noindent where the topological distance or shortest path length $d_{ij}$ is the minimum number of edges that have to be crossed to travel from vertex $i$ to vertex $j$ ($d_{vv} = 0$ by definition). If $i$ and $j$ are not connected, the maximum topological distance in the graph $d_{ij}= N-1$ is used in the sum. Closeness centrality is normalized to $0\le CC_v\le 1$. Following our definition, $CC_v$ is large, when $v$ is topologically close to the rest of the network. One should bear this in mind, because some researchers have used the inverse of our definition \cite{freeman1979csn, GorkaDiss2008}.
 
\subsubsection{Betweenness centrality}

Assume that information travels through the network on shortest paths. There are $\sigma_{ij}$ shortest paths connecting two vertices $i$ and $j$. We then regard a vertex $v$ to be an important mediator for the information transport in the network, if it is traversed by a large number of all existing shortest paths. Mathematically, the betweenness $BC_v$ can be expressed by

\begin{equation}
BC_v = \sum_{i,j\neq v}^N \frac{\sigma_{ij}(v)}{\sigma_{ij}},
\end{equation}

\noindent where $\sigma_{ij}(v)$ gives the number of shortest paths from $i$ to $j$, that include $v$ \cite{freeman1979csn}. Here the contribution of shortest paths is weighted by their respective multiplicity $\sigma_{ij}$.

\subsubsection{Average path length}

The average or characteristic path length $\mathcal{L}$ of a graph is defined as the average topological distance between all pairs of vertices,

\begin{equation}
\mathcal{L} = \frac{1}{{N \choose 2}} \sum_{i<j} d_{ij}.
\end{equation}

\noindent Disconnected pairs of vertices are not included in the average, for a detailed discussion see \cite{Newman2003}.


\section{Constructing climate networks \label{Construction}}

To clarify the physical rational behind our method of climate network construction, we discuss it within the framework of synchronization from dynamical systems theory \cite{UBHD65376642}. In a discretized model of the climate system, dynamical correlations can be envisioned as arising by (partial) synchronization of nonlinear oscillators on the grid that physically form a locally connected network. Even this simple network topology can generate nontrivial spatial patterns of synchronization \cite{Arenas200893,Blasius2005,tonjes:pft}. The same is true for the synchronization of modes of variability in spatially continuous systems as the underlying fields of fluid- and thermodynamics \cite{2002PhR...366....1B}, e.g. SAT. Many measures of synchronization have been proposed and used to infer coupling strength and direction between connected nonlinear oscillators \cite{UBHD65376642,PhysRevLett.76.1804}. The Pearson correlation coefficient \cite{2007NJPh....9..178Z} and the mutual information \cite{Schmidt2008} were successfully employed to retrieve the network topology from the dynamics on the vertices alone.

The concept of synchronization provides a powerful paradigm to guide the enhancement of our understanding of the formation of (nonlinear) teleconnections in the climate system, and to stimulate the development of more advanced measures to detect these effects in measured data \cite{UBHD65376642,2002PhR...366....1B}. We hence propose that research aiming to construct networks from multivariate climatological data should be embedded within the framework of synchronization in complex networks \cite{Arenas200893}.

\subsection{Correlation measures \label{Correlation_measures}}

In the spirit of simplicity facing comparably short time series and desiring consistency with the literature, we choose to first use the standard Pearson correlation coefficient and then cross-check the results by introducing mutual information to climate network construction. The mutual information will allow to investigate nonlinear dynamical relationships (nonlinear teleconnections) that are not fully detectable by using the linear Pearson correlation coefficient \cite{BrockwellDavies2002}. Note that we evaluate both measures at zero lag between time series. In principle, one can calculate a time delayed Pearson correlation (the cross correlation function) and mutual information \cite{kantz2004nts}. This is appropriate when studying climate networks on smaller time scales using data sets with (sub-)diurnal resolution \cite{yamasaki2008,gozol2008,Donner2008nts}. However, in the present work, we intend to study long term structural properties of the climate system on a scale of $\mathcal{O}(10^2)$ years using monthly averaged data. Most physical mechanisms of global information transfer in the SAT field, such as traveling Rossby waves, heat exchange between ocean and atmosphere or the advection of heat by surface currents in the ocean, act on time scales of less than one month. Therefore, it is reasonable to calculate the correlation measures at zero lag between anomaly time series.

\subsubsection{Pearson correlation coefficient}

The parametric empirical Pearson correlation coefficient $R_{ij} =  \left<\hat{a}_i(t) \hat{a}_j(t)\right>_t = R_{ji}$ estimates the strength of a linear relationship between two normalized time series $\hat{a}_i$ and $\hat{a}_j$, given those are normally distributed. It produces spurious results for not normally distributed observables and nonlinear relationships. Consequently it should be used with care when constructing climate networks. The non-parametric Spearman rank order correlation coefficient, that does not depend on the assumption of normally distributed observables, and $R_{ij}$ are found to converge to the same value for nearly all pairs of time series taken from the data sets introduced in Sect. \ref{Data}. The corresponding climate networks hence display close to identical network measures at all topological scales and we conclude, that utilizing the Pearson correlation coefficient to study linear climate networks is statistically justified here.

In contrast to the standard definition of teleconnectivity \cite{wallace1981tgh}, we do not limit our analysis to strongly negative correlations. As in earlier works on climate networks, we use the absolute value of Pearson correlation $P_{ij} = |R_{ij}| = P_{ji}$ to construct climate networks, since both large negative and positive values of Pearson correlation are indicative of a strong linear statistical interdependence. 

\subsubsection{Mutual information}

In climate science, nonlinear measures of statistical interdependence have been successfully applied to uncover strongly nonlinear relationships of climate observables, e.g. the phase coherence between ENSO and the Indian Monsoon \cite{maraun2005epc}. Mutual information from information theory is another nonlinear measure now widely applied in many fields of science, ranging from linguistics \cite{church1990wan} to computational neuroscience \cite{Schmidt2008}. The mutual information $M_{ij}$ can be interpreted as the excess amount of information generated by falsely assuming the two time series $\hat{a}_i$ and $\hat{a}_j$ to be independent, and is able to detect nonlinear relationships \cite{kantz2004nts}. By definition, $M_{ij}$ is large if the two time series are highly linearly (anti)correlated. In contrast, a strongly nonlinear relationship between $\hat{a}_i$ and $\hat{a}_j$ yields large $M_{ij}$, but small $P_{ij}$ (see the upper left quadrant in Fig. \ref{fig:CorrelationLinkDist03}). The mutual information can be estimated using

\begin{equation}
M_{ij} = \sum_{\mu \nu} p_{ij}(\mu, \nu) \log \frac{p_{ij}(\mu, \nu)}{p_i(\mu) p_j(\nu)}, \label{MIFormula}
\end{equation}

\noindent where $p_i(\mu)$ is the probability density function (PDF) of the time series $\hat{a}_i$, and $p_{ij}(\mu, \nu)$ is the joint PDF of a pair $(\hat{a}_i,\hat{a}_j)$. By definition, $M_{ij}$ is symmetric, so that $M_{ij} = M_{ji}$. The standard unit of measurement of mutual information is the bit, if logarithms to base 2 are used.

We use a simple histogram approach with equally sized bins for all pairs $\{i,j\}$ to estimate the probability densities. Because the estimator (Eq. \ref{MIFormula}) is known to depend on bin size and partitioning \cite{Schwarz1993,hegger-1999-9,papana2008emi}, we use an identical partitioning for all $\{i,j\}$ to guarantee an optimal comparability of the $M_{ij}$. We select a bin number of 64, that meets the Cochran criterion of at least 5 samples per bin for a typical time series length of $\mathcal{O}(10^3)$. The basic algorithm applied here is computationally much less expensive than more advanced methods proposed in the literature \cite{papana2008emi,PhysRevE.69.066138}, which is an important advantage when dealing with up to $\mathcal{O}(10^8)$ pairs in a global climate network. Our algorithm is feasible, since the application to network construction requires only the correct estimation of relative differences of $M_{ij}$ between all pairs of time series. In other words, in our application systematic under- or overestimation of mutual information is not a problem, as long as the error stays approximately constant across all pairs.

\subsection{Obtaining the network adjacency matrix \label{Adjacency_matrix}}

We now construct the climate network by thresholding the correlation measure matrix $C_{ij}$ ($C_{ij} = P_{ij}$ or $C_{ij} = M_{ij}$), i.e. only pairs of vertices $\{i,j\}$ that satisfy $C_{ij}>\tau$ are regarded as linked. By definition $C_{ij} \geq 0$, $\forall \{i,j\}$ (see Sect. \ref{Correlation_measures}). Using the Heaviside function $\Theta(x)$, the adjacency matrix $A_{ij}$ of the climate network is then given by 

\begin{equation}
A_{ij} = \Theta \left(C_{ij} - \tau \right) - \delta_{ij},
\end{equation}

\noindent where $\delta_{ij}$ is the Kronecker delta. Note that $A_{ij}$ inherits its symmetry from $C_{ij}$ and the resulting climate network is an undirected and unweighted simple graph.

\subsection{Choosing the threshold \label{Choosing_threshold}}

\begin{figure}
\centering

\subfigure[]{
\includegraphics[width=0.48 \textwidth]{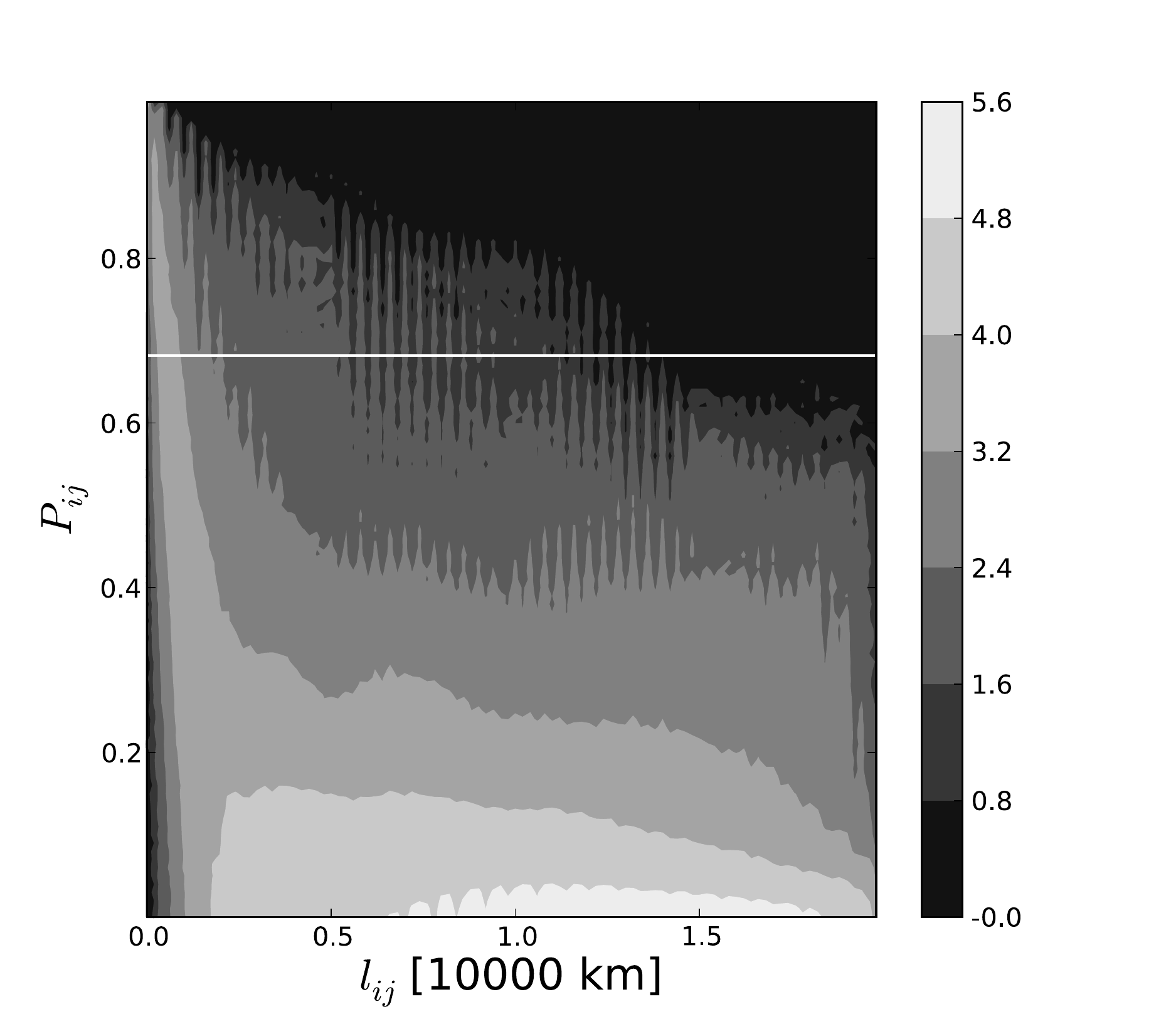}
\label{fig:CorrelationLinkDist01}
}
\subfigure[]{
\includegraphics[width=0.48 \textwidth]{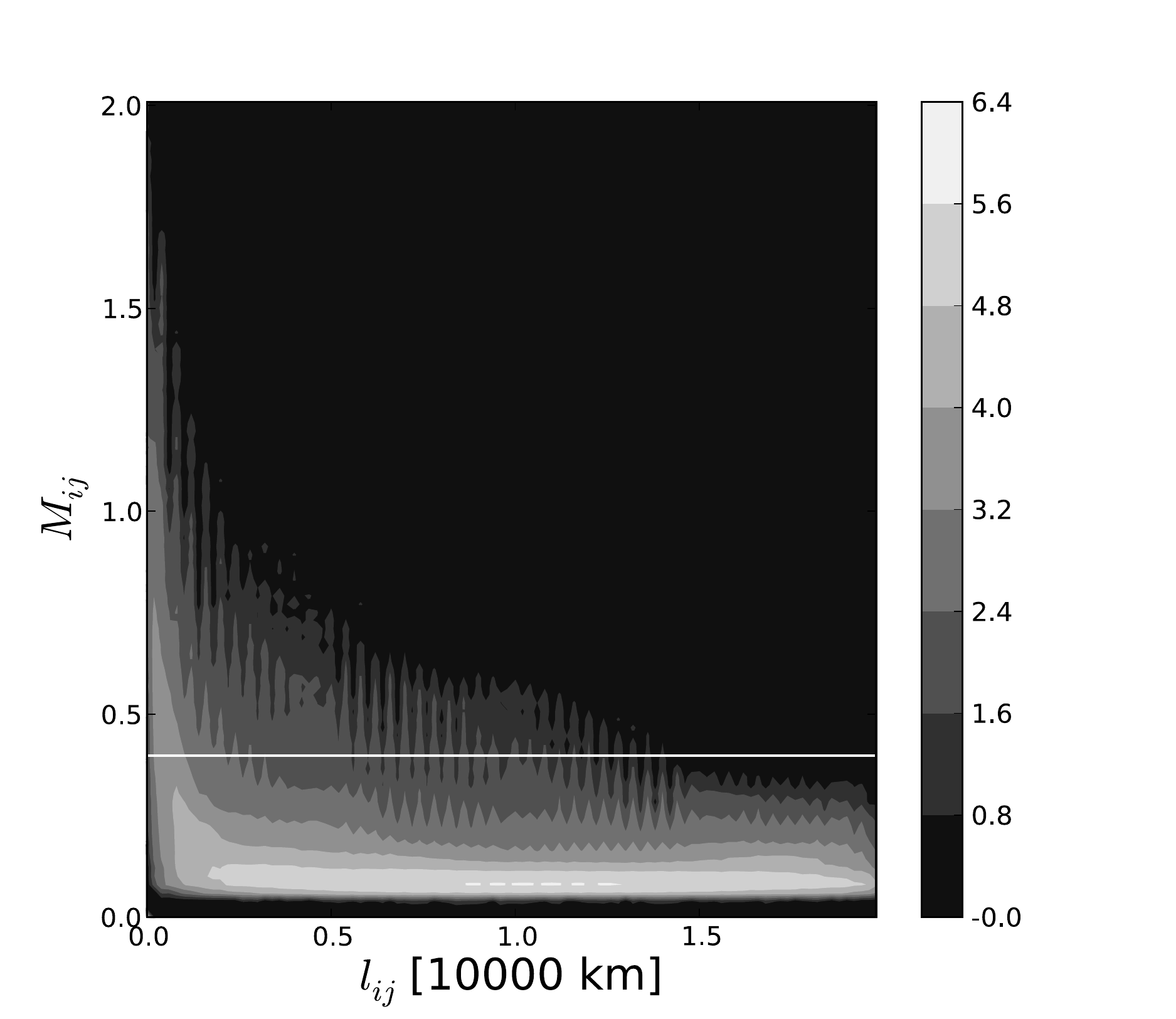}
\label{fig:CorrelationLinkDist02}
}
\subfigure[]{
\includegraphics[width=0.48 \textwidth]{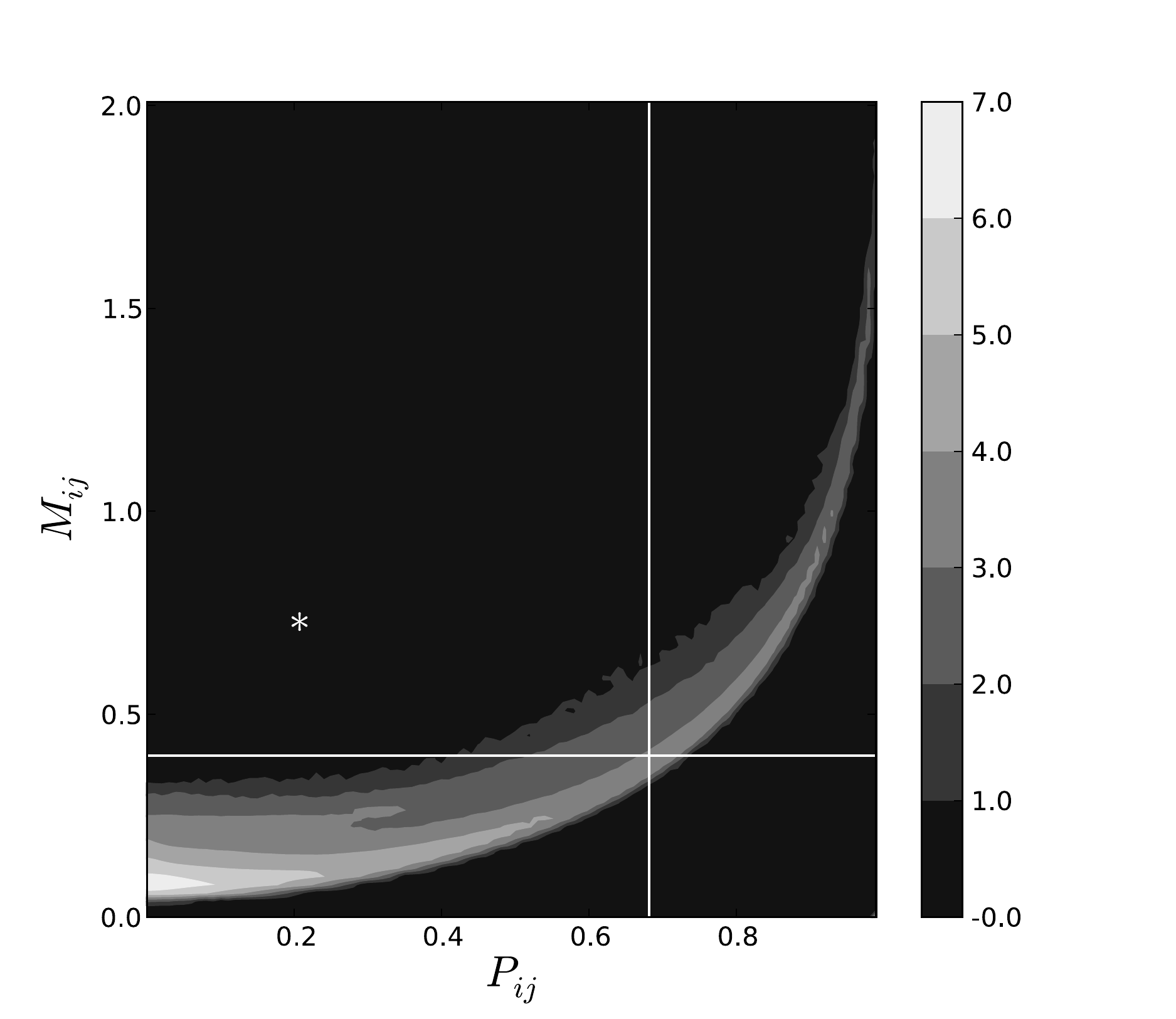}
\label{fig:CorrelationLinkDist03}
}

\caption{\label{fig:CorrelationLinkDist} (a,b) Frequency plot in the space of correlation measure $C_{ij}$ and edge distance $l_{ij} = R_{earth} \arccos ( \sin(\lambda_i)\sin(\lambda_j) + \cos(\lambda_i)\cos(\lambda_j)\cos(\phi_i-\phi_j) )$ for all $N(N-1) / 2 =23,228,928$ pairs of time series in the global HadCM3 SAT data set. The apparent oscillations with edge distance are an artifact of the finite spatial resolution of the underlying grid. (c) Frequency plot in the space of Pearson correlation $P_{ij}$ and mutual information $M_{ij}$. All plots are based on 2D-histograms with $10^4$ equally sized rectangular bins. The color bars indicate the common logarithm of frequency. Vertical and horizontal lines mark the thresholds corresponding to edge density $\rho=0.005$ for $P_{ij}$ and $M_{ij}$ (Fig. \ref{fig:CorrelationDist}). The asterisk in (c) delineates the quadrant containing edges that exist in the mutual information, but not in the Pearson correlation network of $\rho=0.005$, and hence are candidates for strongly nonlinear connections.}
\end{figure}

The last but nontrivial step in climate network construction is the selection of a threshold $\tau$, above which we consider a pair of vertices to be connected. From a statistical point of view it is desirable to only maintain connections that are statistically significant with respect to some reasonable test and reject those not meeting this criterion. Classical significance tests and randomization experiments have been used to assess the value of $\tau$ for climate networks constructed using the Pearson correlation coefficient \cite{tsonis2004acn,tsonis2006,tsonis2008tap}. We build on these results testing against randomly shuffled time series, Fourier surrogates and twin surrogates \cite{thiel2006tst}. Twin surrogates correspond to the null hypothesis of trajectories with random initial conditions on the attractor of the original time series and are found to give the strictest bounds on the significance of network connections detected using Pearson correlation and mutual information.

\subsubsection{On the role of teleconnections}

From the perspective of complex network theory, we intend to uncover interesting structures in the topology of the climate network. Different features of the underlying correlation measure matrix $C_{ij}$ will be revealed at different thresholds $\tau$. Consequently, the choice of $\tau$ has to reflect a trade-off between the statistical significance of connections and the richness of network structures unveiled. For example, note the potentially interesting long distance edges with high Pearson correlation and mutual information at edge distance $l \gtrsim 15000 \textrm{km}$ in the global HadCM3 SAT data set (Fig. \ref{fig:CorrelationLinkDist01} and \ref{fig:CorrelationLinkDist02}). They will only be included in the climate network, if the threshold $\tau \lesssim 0.65$ for the Pearson correlation network, or $\tau \lesssim 0.3$ in the case of the mutual information network. Long distance edges with high correlation measure or teleconnections are responsible for all interesting and non-trivial features of climate networks, such as small-world behavior, super-nodes or betweenness structures. Without them serving as spatial short cuts in the network, only the locally connected underlying grid remains. Ergo the inclusion of teleconnections must be a necessary criterion in the choice of the threshold in order to obtain interesting results in climate network analysis.

\subsubsection{Dependence of network measures on edge density}

\begin{figure}
\centering
\subfigure[]{
\includegraphics[width=0.48 \textwidth]{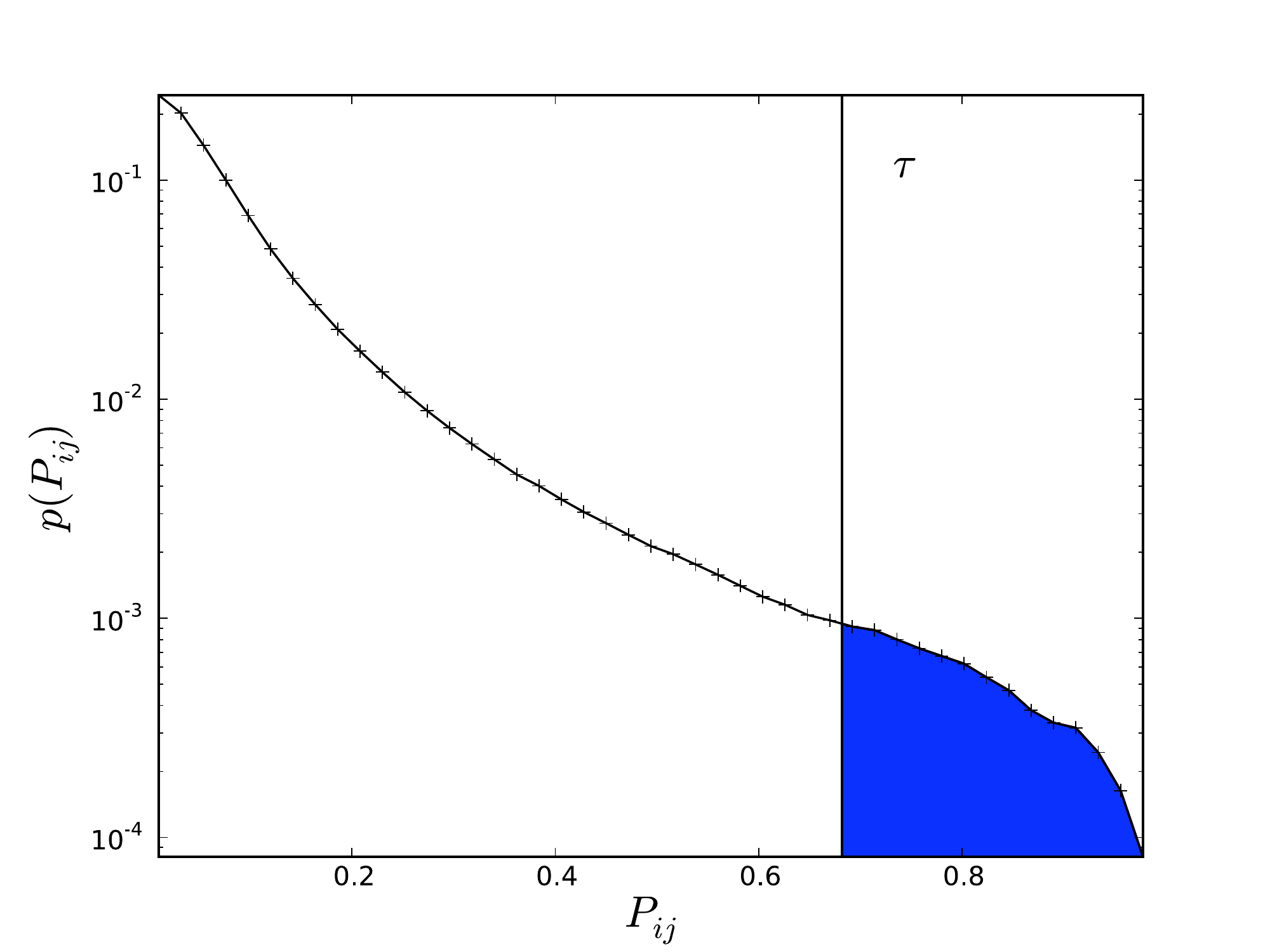}
\label{fig:CorrelationDist01}
}
\subfigure[]{
\includegraphics[width=0.48 \textwidth]{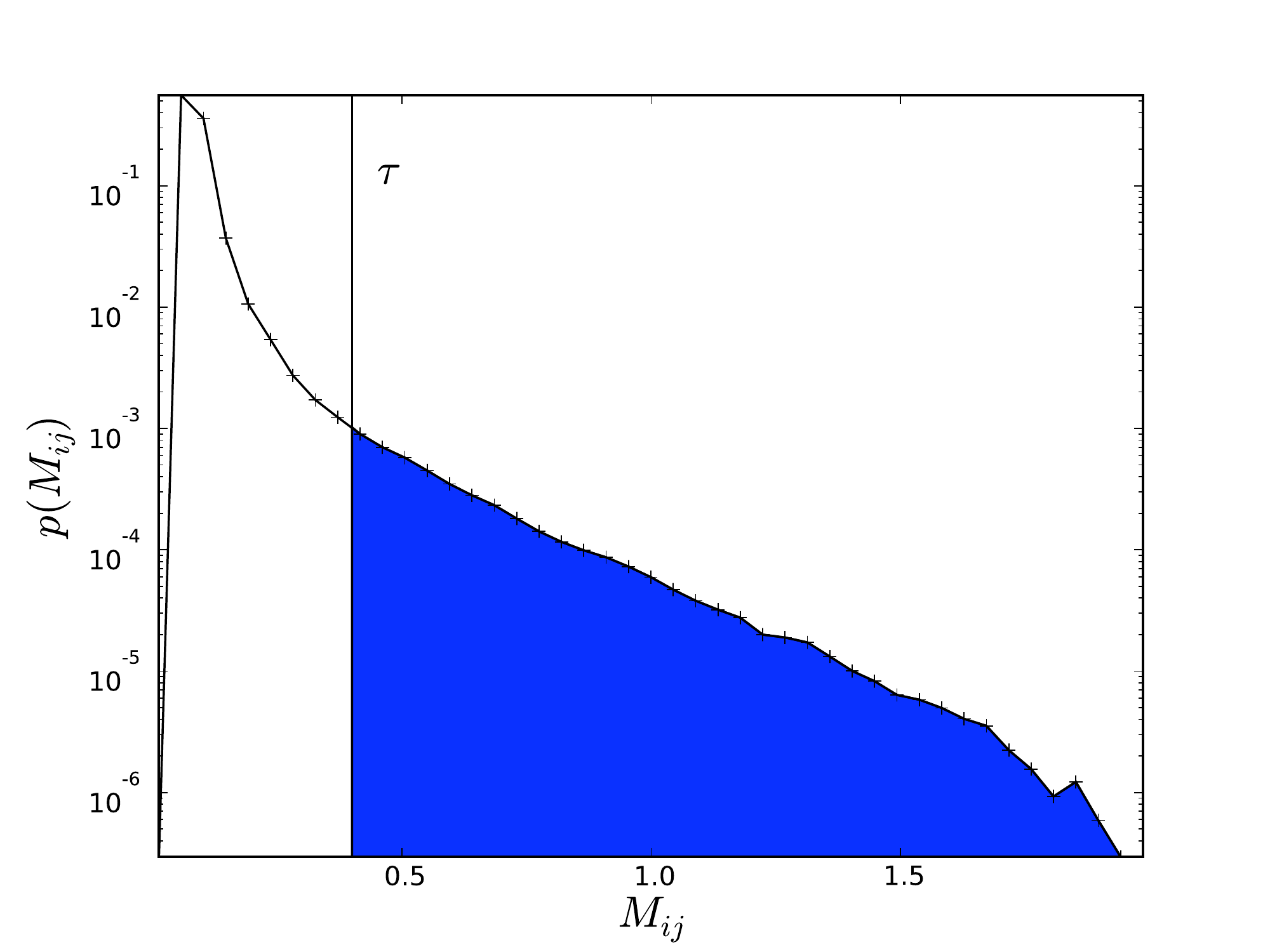}
\label{fig:CorrelationDist02}
}
\caption{\label{fig:CorrelationDist} PDF $p(C)$ of the correlation measure matrices $C_{ij}$ for the HadCM3 SAT data set. The vertical line indicates the threshold $\tau$ yielding an edge density $\rho(\tau) = 0.005$, that is equal to the shaded area. (a) Pearson correlation, $\tau = 0.682$, b) mutual information, $\tau = 0.398$.}
\end{figure}

\begin{figure}
\centering

\subfigure[]{
\includegraphics[width=0.48 \textwidth]{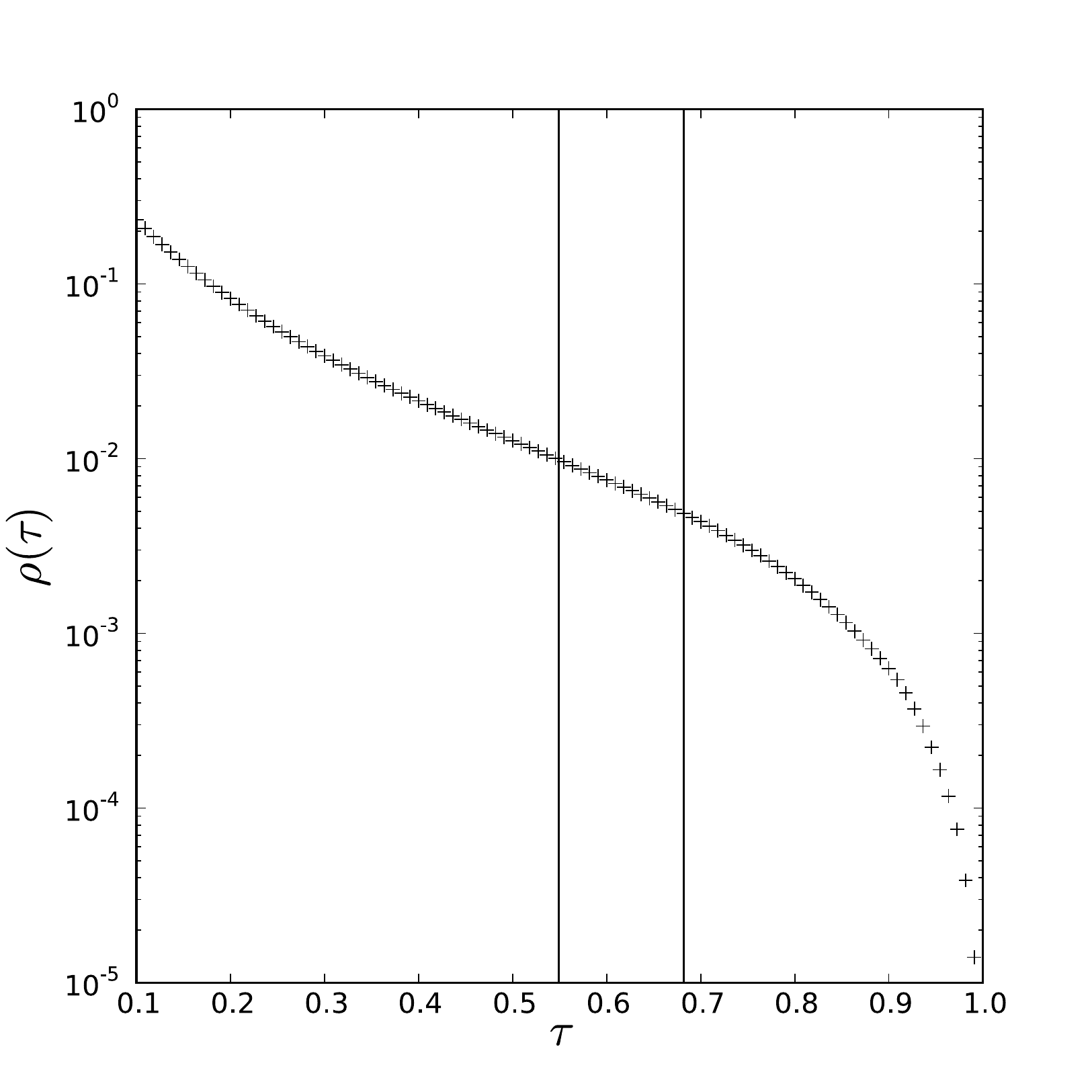}
\label{fig:ThresholdDependence01}
}
\subfigure[]{
\includegraphics[width=0.48 \textwidth]{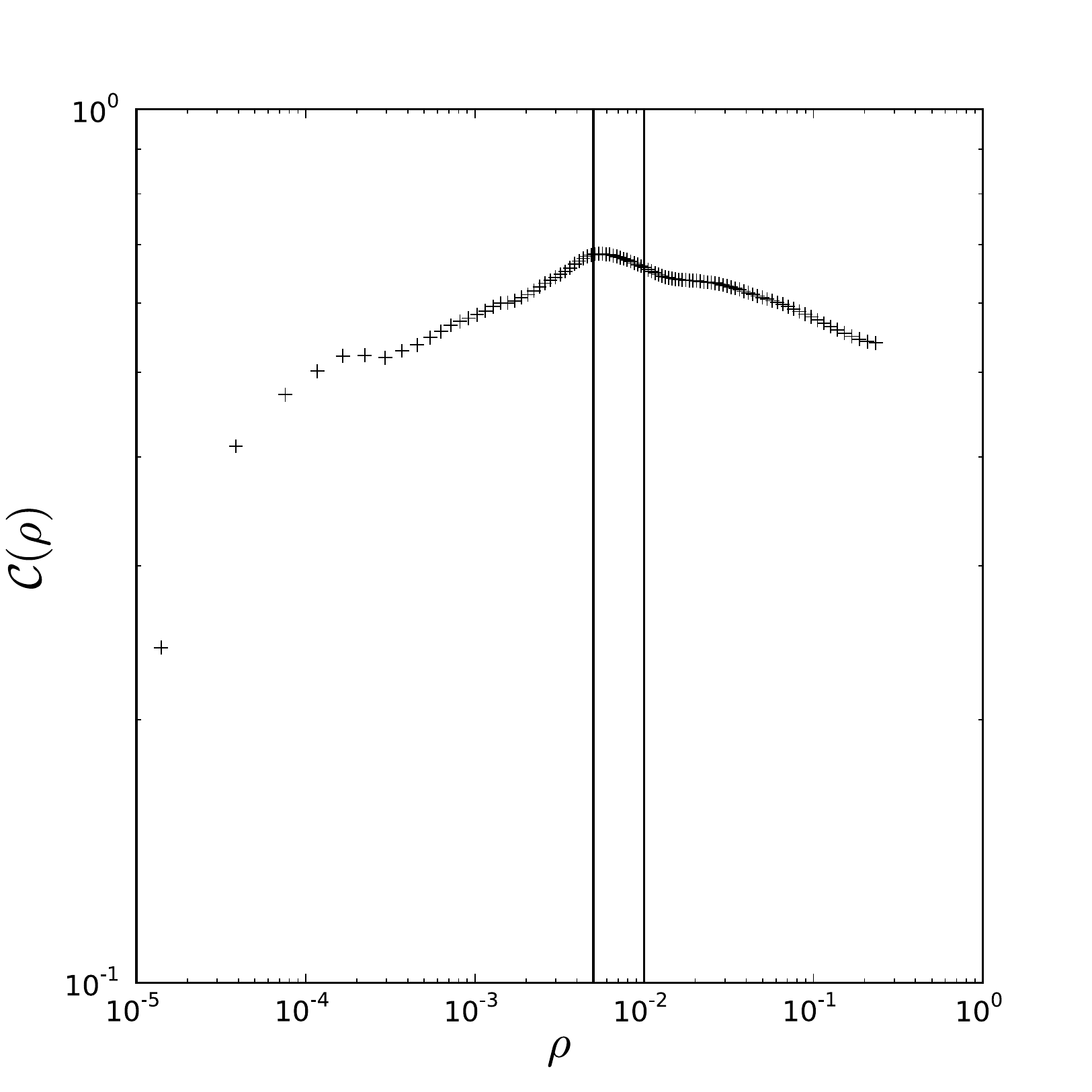}
\label{fig:ThresholdDependence02}
}
\subfigure[]{
\includegraphics[width=0.48 \textwidth]{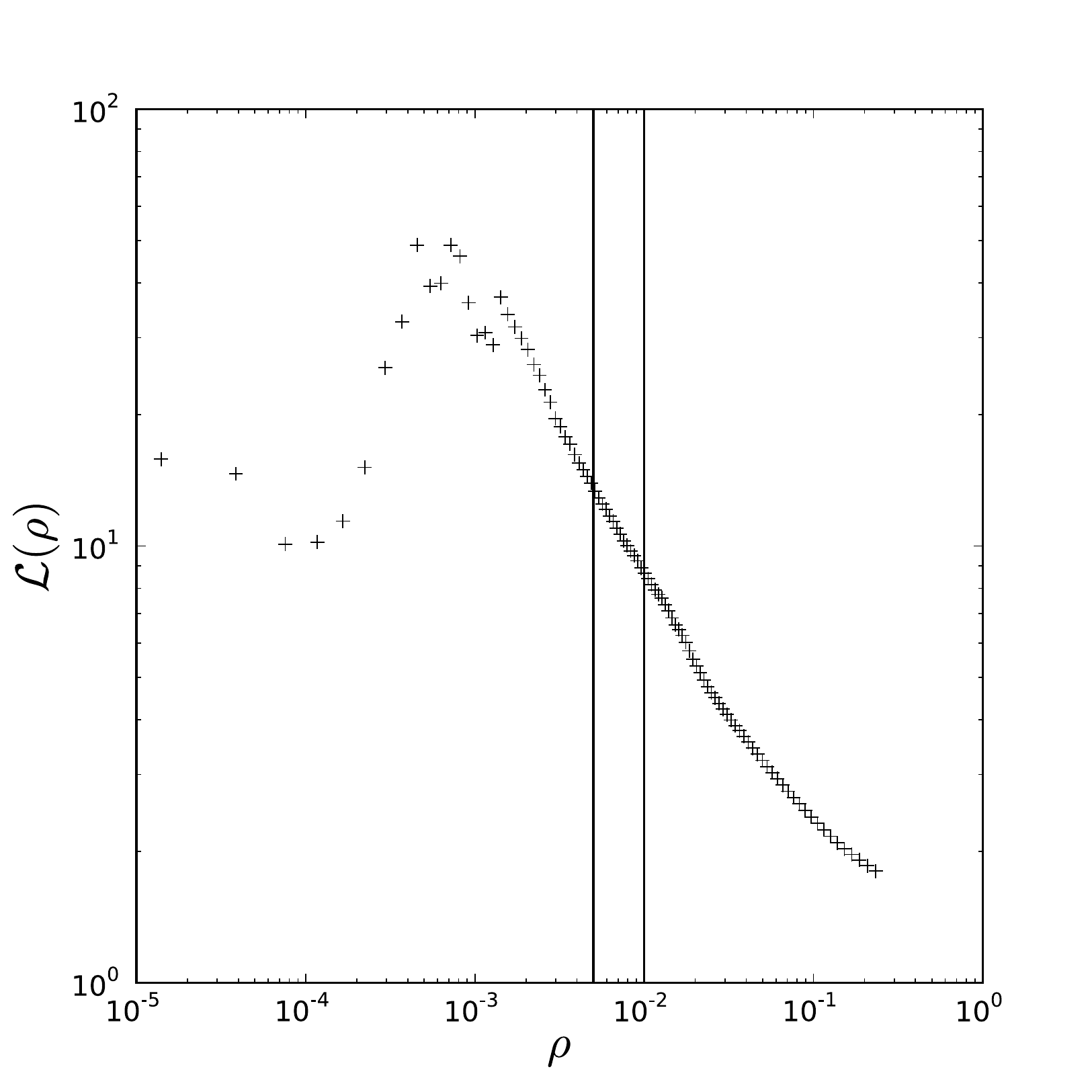}
\label{fig:ThresholdDependence03}
}
\subfigure[]{
\includegraphics[width=0.48 \textwidth]{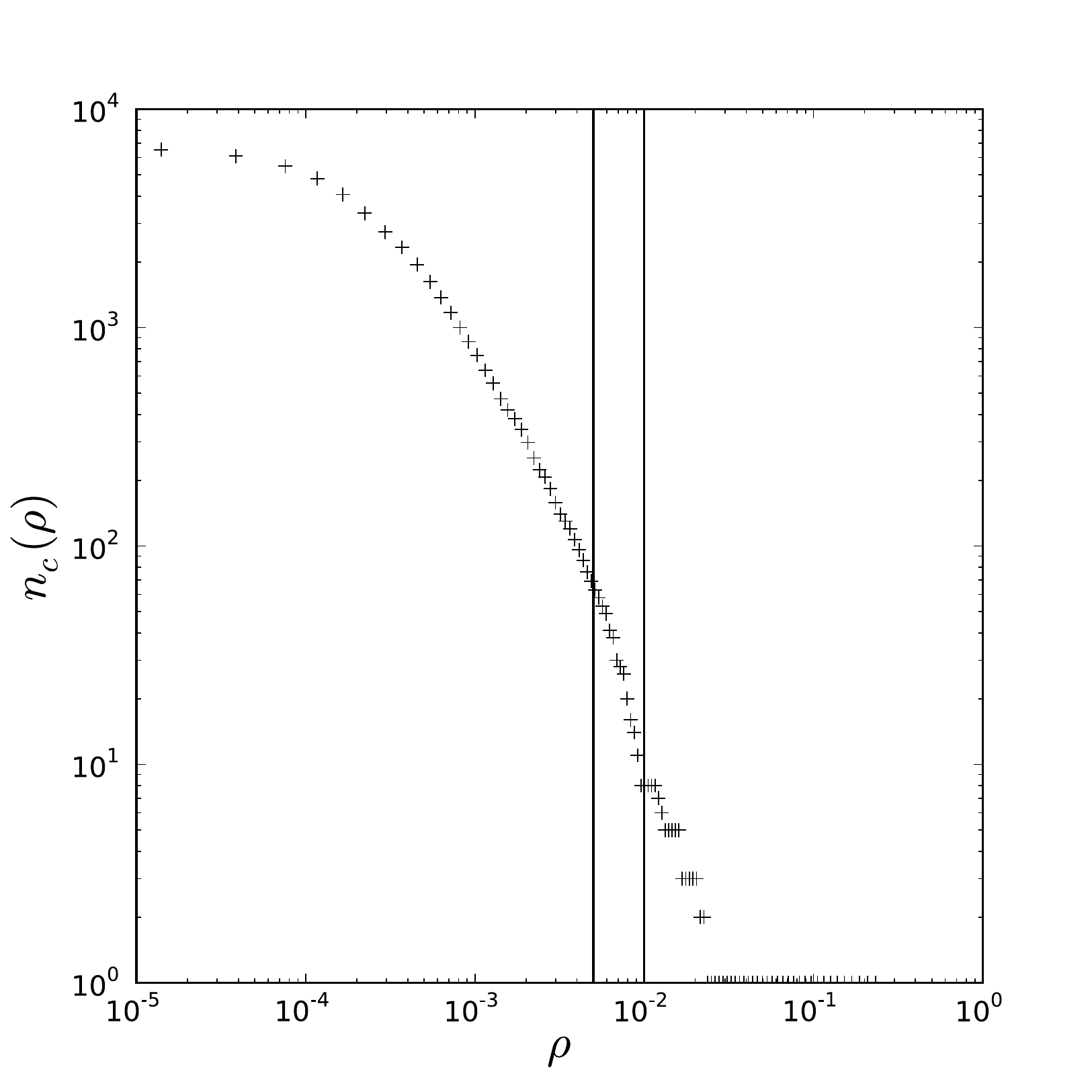}
\label{fig:ThresholdDependence04}
}
\subfigure[]{
\includegraphics[width=0.48 \textwidth]{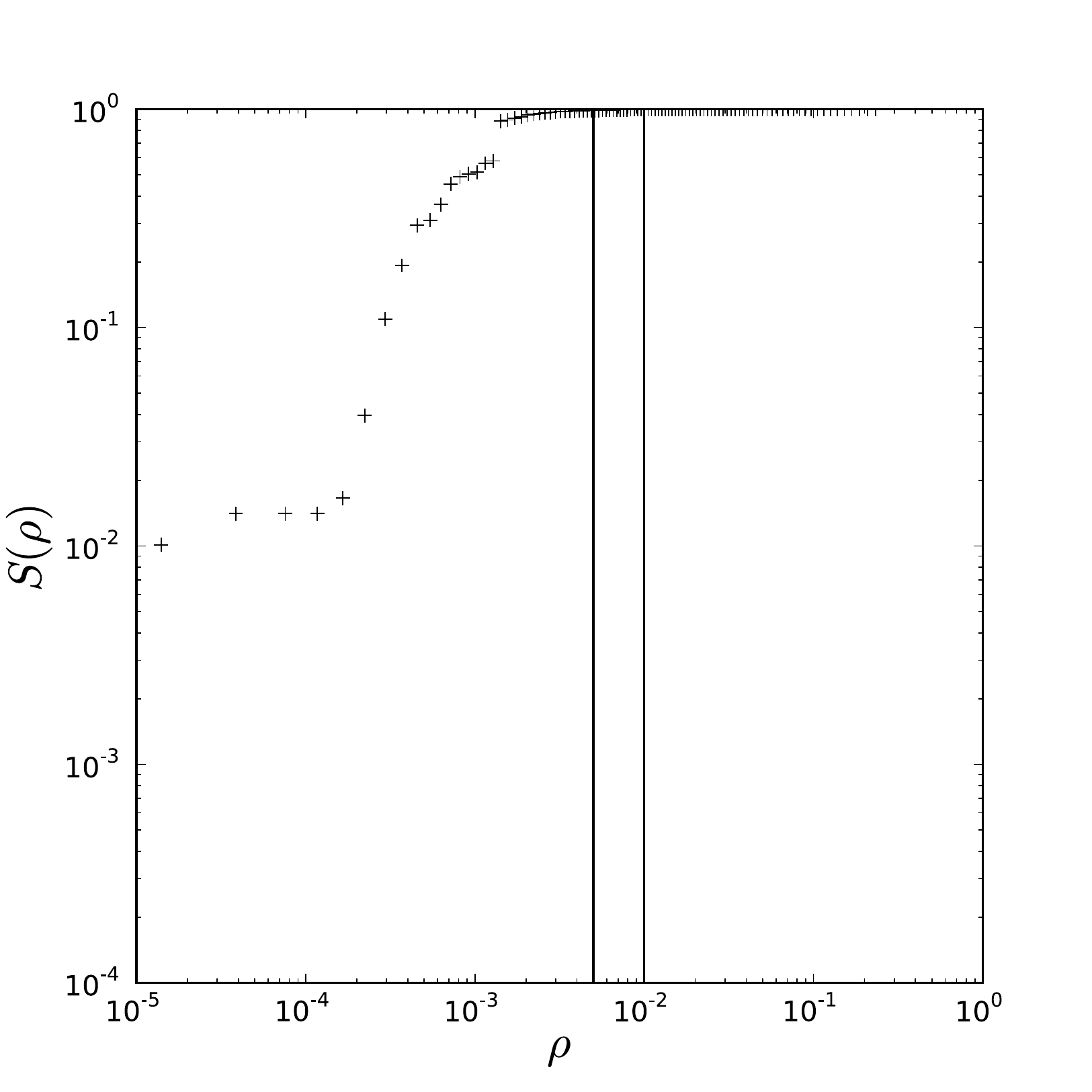}
\label{fig:ThresholdDependence05}
}
\subfigure[]{
\includegraphics[width=0.48 \textwidth]{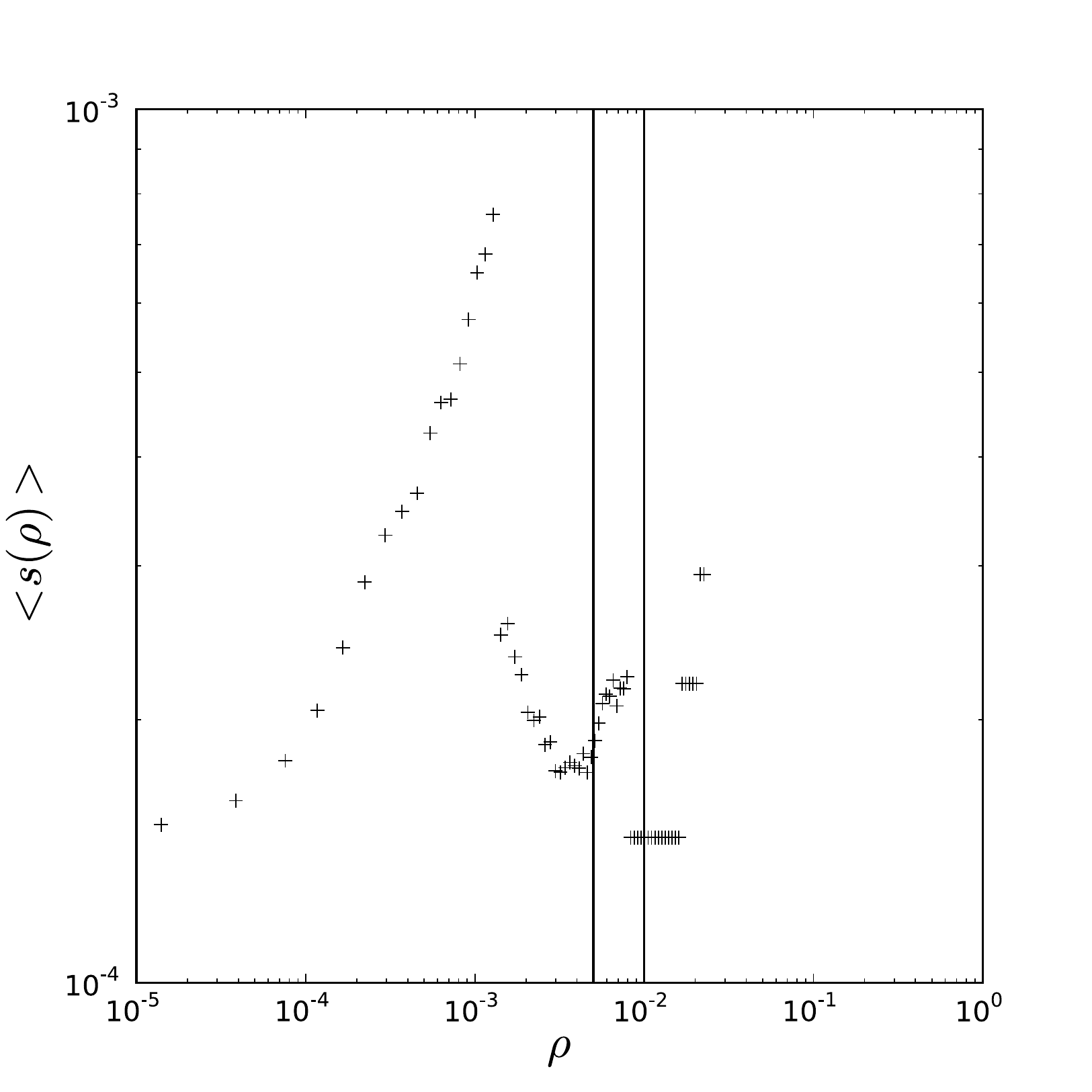}
\label{fig:ThresholdDependence06}
}

\caption{\label{fig:ThresholdDependence}Network measures as a function of threshold and edge density for global HadCM3 SAT networks constructed using Pearson correlation. (a) Threshold dependence of edge density $\rho(\tau)$, (b) edge density dependence of clustering coefficient $\mathcal{C}(\rho)$ and (c) average path length $\mathcal{L}(\rho)$. (d) Edge density dependence of the number of components $n_c(\rho)$, (e) giant component size $S(\rho)$ and (f) average non-giant component size $\left<s(\rho)\right>$. The vertical lines indicate edge densities of $\rho=0.005$ and $\rho=0.01$ and corresponding thresholds.}
\end{figure}

Systematic studies show a smooth dependence of most climate network measures on $\tau$ in the range of edge densities considered in this work. This implies that small uncertainties in the choice of the threshold will not lead to strongly deviating results within the complex network framework. Here we discuss the threshold dependence of edge density $\rho(\tau)$, and the edge density dependence of clustering coefficient $\mathcal{C}(\rho)$, average path length $\mathcal{L}(\rho)$, number of components $n_c(\rho)$, relative giant component size $S(\rho)$ and average relative non-giant component size $\left<s(\rho)\right>$  (Fig. \ref{fig:ThresholdDependence}). Here a component constitutes a maximally connected subset of vertices of the network, i.e. a connected subset of vertices that is not reachable from any other vertex in the network. The term giant component is usually reserved for the largest component containing nearly all of the vertices in the network \cite{Newman2003}. $S(\rho)$ in turn always measures the relative size of the largest component, even if its size becomes comparable to that of other components. 

The edge density $\rho(\tau)$ decays approximately exponentially due to the shape of the PDF of the absolute value of the correlation measure $p(C)$ (in the following we abbreviate $C_{ij}$ by $C$),

\begin{equation}
\rho(\tau) = \int_{\tau}^{\infty}dC p(C).
\end{equation}

\noindent Note that $\rho(\tau)$ is a monotonic decreasing function of $\tau$. Correlation measure distributions found empirically from climate data generally have a connected support (Fig. \ref{fig:CorrelationDist}), so that $\rho(\tau)$ is strictly monotonic decreasing and induces a one to one correspondence between threshold $\tau$ and edge density $\rho$ (Fig. \ref{fig:ThresholdDependence01}).

The clustering coefficient $\mathcal{C}$ is found to stay approximately constant at intermediate values of $\rho$ and decays to zero for small $\rho$ (Fig. \ref{fig:ThresholdDependence02}), when the network decomposes into a larger number $n_c$ (Fig. \ref{fig:ThresholdDependence04}) of smaller components (Fig. \ref{fig:ThresholdDependence05} and \ref{fig:ThresholdDependence06}). The average path length $\mathcal{L}$ decays approximately as a power law with growing $\rho$ and has discontinuities at edge densities $\rho_\mu$, where $\tau_\mu = \tau(\rho_\mu)$ equals the correlation measure $C_{ij}$ of edges $\{i,j\}$ with a high edge betweenness centrality \cite{Newman2003}, i.e. that lie on many shortest paths between pairs of vertices (Fig. \ref{fig:ThresholdDependence03}). When $\tau \geq \tau_\mu$, these shortest paths become considerably longer and components might decouple from the network's giant component. This effect leads to a decrease of $\mathcal{L}$ for small $\rho$ since the network decomposes into smaller disconnected components (Fig. \ref{fig:ThresholdDependence06}) and path lengths are measured only within the components. The formation of a giant component encompassing nearly all vertices at $\overline{\rho} \approx 0.0012$, where the giant component size increases from $S \approx 0.5$ to $S \approx 1$ (Fig. \ref{fig:ThresholdDependence05}), goes along with discontinuities of $\mathcal{L}$ and $\left<s(\rho)\right>$. Note that all vertices have joined the giant component at $\rho \approx 0.020$ for the HadCM3 SAT Pearson correlation network (Fig. \ref{fig:ThresholdDependence04}) and at $\rho \approx 0.028$ for the corresponding mutual information network (not shown here).

At all edge densities considered in Sect. \ref{Results} the giant component size is of $\mathcal{O}(1)$. The influence of the non-giant components on measures such as average path length and closeness centrality is therefore negligible in the regime studied here, since larger deviations are only expected for $\rho<\overline{\rho}$. This range of edge densities in turn is not relevant for the conclusions drawn from the comparison presented in Sect. \ref{Results}. To study this regime of very small edge densities in detail, measures more robust to disconnected components such as the local efficiency (related to closeness centrality) and global efficiency (related to average path length) should be considered \cite{Newman2003}. We chose the definitions given in Sect. \ref{Network_measures} to maintain consistency with the existing literature on climate networks.

\subsubsection{Pragmatic choice of $\tau$}

We think that the problem of selecting exactly the right threshold is not as severe as might be thought. Climate network analysis deals with topological properties of correlation measure matrices and aims at gaining new insights heeding this paradigm. In the climate system, it is furthermore not immediately evident which physical entities should take the role of vertices and edges in a complex network. This constitutes the main conceptional difference between our method and attempts of recovering an unknown physically existent network structure from vertex dynamics as in the study of the brain \cite{2007NJPh....9..178Z,Schmidt2008,2006RvMP...78.1213R,2006PhRvL..97w8103Z,Peter2008}, where one can argue that a more natural identification of neurons and axons with the vertices and edges of a neural network exists. It is known that in the classical local description of geophysical fluid dynamics of atmosphere and oceans, i.e. the Navier-Stokes equations combined with thermodynamic equations, the network of physical interaction has the structure of a regular grid \cite{vallis2006aao}. In a discretized model, the dynamics at each grid point is only coupled to the grid points in the immediate neighborhood. The complex topology observed in climate networks should therefore be treated as a manifestation of structure formation, that allows for uncertainties in the choice of parameters such as $\tau$.

In the spirit of the ideas elaborated in the above paragraphs, we choose to fix the edge density $\rho$ when comparing the properties of climate networks generated using different correlation measures. This will result in different thresholds $\tau$, because the empirical correlation measure distribution $p(C)$ clearly differs between linear Pearson correlation and nonlinear mutual information (Fig. \ref{fig:CorrelationDist}). The selection of $\rho$ is in each case guided by the principle of balancing between structural richness and statistical significance outlined above.


\section{Results \label{Results}}

After having introduced our methodology for climate network construction, we proceed to the main aim of this study:  A comparison of networks generated using the linear Pearson correlation coefficient and the nonlinear mutual information on local, mesoscopic and global topological scales. The edge density $\rho$ is varied between $\rho_{min}=0$ and $\rho_{max}=0.1$ in equally sized steps. Recall, that small edge densities correspond to high thresholds (Sec. \ref{Choosing_threshold}). For increasing edge density, edges with decreasing correlation measure are added to the network. Consequently, climate networks with a very high edge density $\rho\ge0.1$ are not expected to contain meaningful information for climate data analysis, because they contain many connections that are not statistically significant, i.e. that are much more likely to arise by chance. For example, Tsonis et al. use the Pearson correlation coefficient and a threshold of $\tau=0.5$ in all of their works \cite{tsonis2004acn,tsonis2006,tsonis2008jclim,tsonis2008tap}, which corresponds to an edge density of $\rho \approx 0.01$ for the global HadCM3 SAT data set analyzed here. They report that according to the Student's t test, a value of $P_{ij}=0.5$ is statistically significant above the 99\% level. In our recent work, we use an edge density of $\rho=0.005$ \cite{Donges2008}. This larger threshold corresponds to an even higher significance level, because it is less likely to be exceeded by the correlation measures calculated from pairs of one original and one surrogate time series.

We compare the properties of the complex networks obtained at each edge density level on local, mesoscopic and global topological scales. We enable a qualitative discussion of similarity by plotting the fields of area weighted connectivity (Fig. \ref{fig:ComparisonAWC}), local clustering coefficient (Fig. \ref{fig:ComparisonC}), closeness (Fig. \ref{fig:ComparisonCC}) and betweenness centrality (Fig. \ref{fig:ComparisonBC}) on a world map at fixed edge density $\rho=0.005$. The local deviations of these fields calculated for Pearson correlation and mutual information climate networks are highlighted by normalized difference fields (Fig. \ref{fig:DifferenceFields}). For a quantitative comparison at all edge densities considered, we calculated the Spearman rank order correlation coefficient or Spearman's Rho $r_s(\rho)$ of the corresponding fields taken from the Pearson correlation and mutual information networks (Fig. \ref{fig:HadCM3_SAT_Comp04} and Fig. \ref{fig:Reanalysis_SAT_Comp04}). We chose to use the Spearman's Rho instead of the Pearson correlation coefficient for this task, because it is known to be more reliable when applied to data with non-Gaussian PDF. This is an important property, considering that some of the fields we are interested in have a highly non-normal frequency distribution (Sect. \ref{Local_comparison} and Sect. \ref{Global_comparison}). Furthermore at each edge density step, we consider the Hamming distance between the networks on the local topological scale, whereas on the mesoscopic and global scale we compare global clustering coefficient and average path length.

In the following we will illustrate the comparison for the HadCM3 SAT data set in detail (Sect. \ref{Local_comparison}, \ref{Mesoscopic_comparison}, \ref{Global_comparison} and Fig. \ref{fig:ComparisonAWC}, \ref{fig:ComparisonC}, \ref{fig:ComparisonCC}, \ref{fig:ComparisonBC}, \ref{fig:DifferenceFields}, \ref{fig:HadCM3_SAT_Comp}). Only the quantitative comparison is presented for the NCEP/NCAR reanalysis SAT data set (Fig. \ref{fig:Reanalysis_SAT_Comp}), since we are lead to the same conclusions as for the model data set. Finally we present climatological interpretations of the observed network structures (Sect. \ref{Climatological_Interpretation}).

\begin{table}
\caption{\label{ComparisonTable}Spearman's Rho $r_s(\rho)$ of area weighted connectivity (AWC), local clustering coeffcient (C), closeness centrality (CC) and betweenness centrality (BC) fields and Hamming distances $H(\rho)$ and $H^R(\rho)$ calculated from Pearson correlation and mutual information networks at edge densities $\rho=0.005$ and $\rho=0.01$ for the global HadCM3 SAT data set.}
\center
\begin{tabular}{l|c|r}
 & $\rho=0.005$ & $\rho=0.01$\\
 \hline
 \hline
$r_s^{AWC}$ & 0.95 & 0.88 \\
$r_s^{C}$ & 0.80 & 0.81 \\
$r_s^{CC}$ & 0.98 & 0.95 \\
$r_s^{BC}$ & 0.70 & 0.59 \\
$H$ & 0.001 & 0.003 \\
$H^R$ & 0.010 &  0.02 \\
\end{tabular}
\end{table}

\subsection{Local comparison \label{Local_comparison}}

\begin{figure}[tp]
\centering

\subfigure[]{
\includegraphics[width=\textwidth, viewport=35 85 715 510, clip]{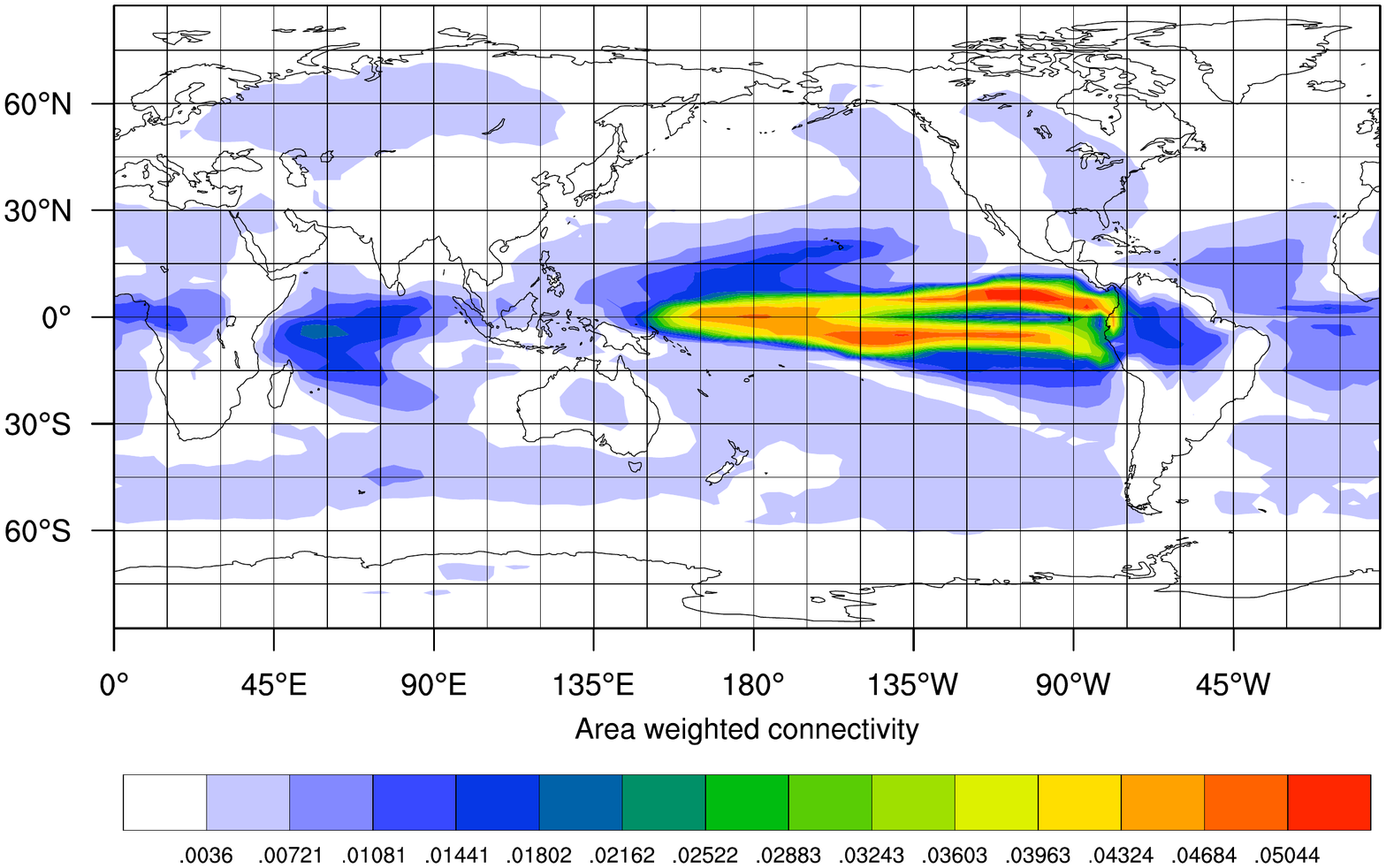}
\label{fig:ComparisonAWCSub1}
}
\subfigure[]{
\includegraphics[width=\textwidth, viewport=35 85 715 510, clip]{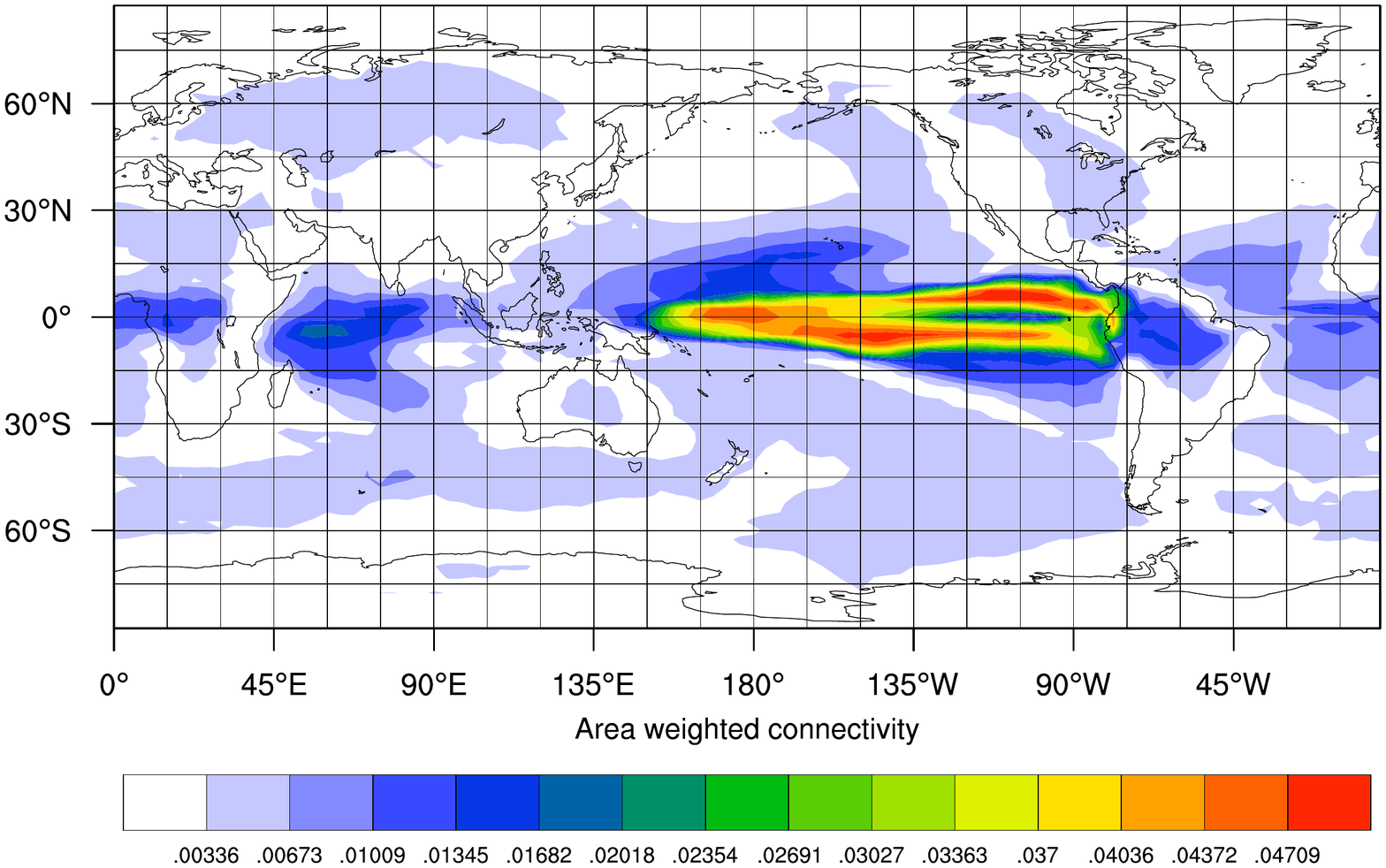}
\label{fig:ComparisonAWCSub2}
}

\caption{\label{fig:ComparisonAWC}Area weighted connectivity fields for global HadCM3 SAT networks at $\rho = 0.005$ (linear color scale) obtained using a) Pearson correlation, b) mutual information. The rank order correlation between the two fields is $r_s^{AWC}(0.005) = 0.95$.}
\end{figure}

On the local topological scale, we find that Pearson correlation and mutual information climate networks are very similar at low edge densities. At $\rho=0.005$, the area weighted connectivity (Fig. \ref{fig:ComparisonAWC}) field shows only small deviations by visual inspection, that are most pronounced in the tropics (Fig. \ref{fig:DifferenceFields01}). The rank order correlation coefficient $r_s^{AWC}$ reaches a maximum between $\rho=0.005$ and $\rho=0.01$ and decays for larger edge densities (Fig. \ref{fig:HadCM3_SAT_Comp04}). We obtain high values for $\rho=0.005$ and $\rho=0.01$ (Table \ref{ComparisonTable}). Note that for the climate networks studied, area weighted connectivity has a fat tailed PDF \cite{tsonis2006}.

The Hamming distance $H(\rho)$ is always smaller than the expected distance $H^R(\rho)$ of two random networks at edge density $\rho$ (Fig. \ref{fig:HadCM3_SAT_Comp01}). It is notable, that $H(\rho)$ seems to go to zero tangentially to the $\rho$-axis, i.e. $H'(\rho)|_{\rho=0} \approx 0$, whereas $H^R(\rho)|_{\rho=0}=2$. Therefore most of the edges with the highest Pearson correlation and mutual information values must coincide. From analytical considerations and Monte-Carlo simulations we find that the standard deviation of the PDF of Hamming distance between the two random networks is of $\mathcal{O}(N^{-1})$ for $N \gg 1$. This means that the expected deviations from the mean $H^R(\rho)$ are of $\mathcal{O}(10^{-4})$ for the climate networks considered here. The difference between measured Hamming distance and $H^R(\rho)$ is by one order of magnitude larger than these expected deviations (Table \ref{ComparisonTable}). We hence conclude that the observed similarity of Pearson correlation and mutual information networks can be considered statistically significant, with respect to the null hypothesis of random networks of the same size $N$, at all edge densities considered. Particularly, the results elaborated in this section show, that at the edge densities used in earlier works on climate networks \cite{tsonis2004acn,tsonis2006,tsonis2008jclim,tsonis2008tap,Donges2008}, Pearson correlation and mutual information give very similar results on the local topological scale.

\subsection{Mesoscopic comparison \label{Mesoscopic_comparison}}

\begin{figure}[tp]
\centering

\subfigure[]{
\includegraphics[width=\textwidth, viewport=35 85 715 510, clip]{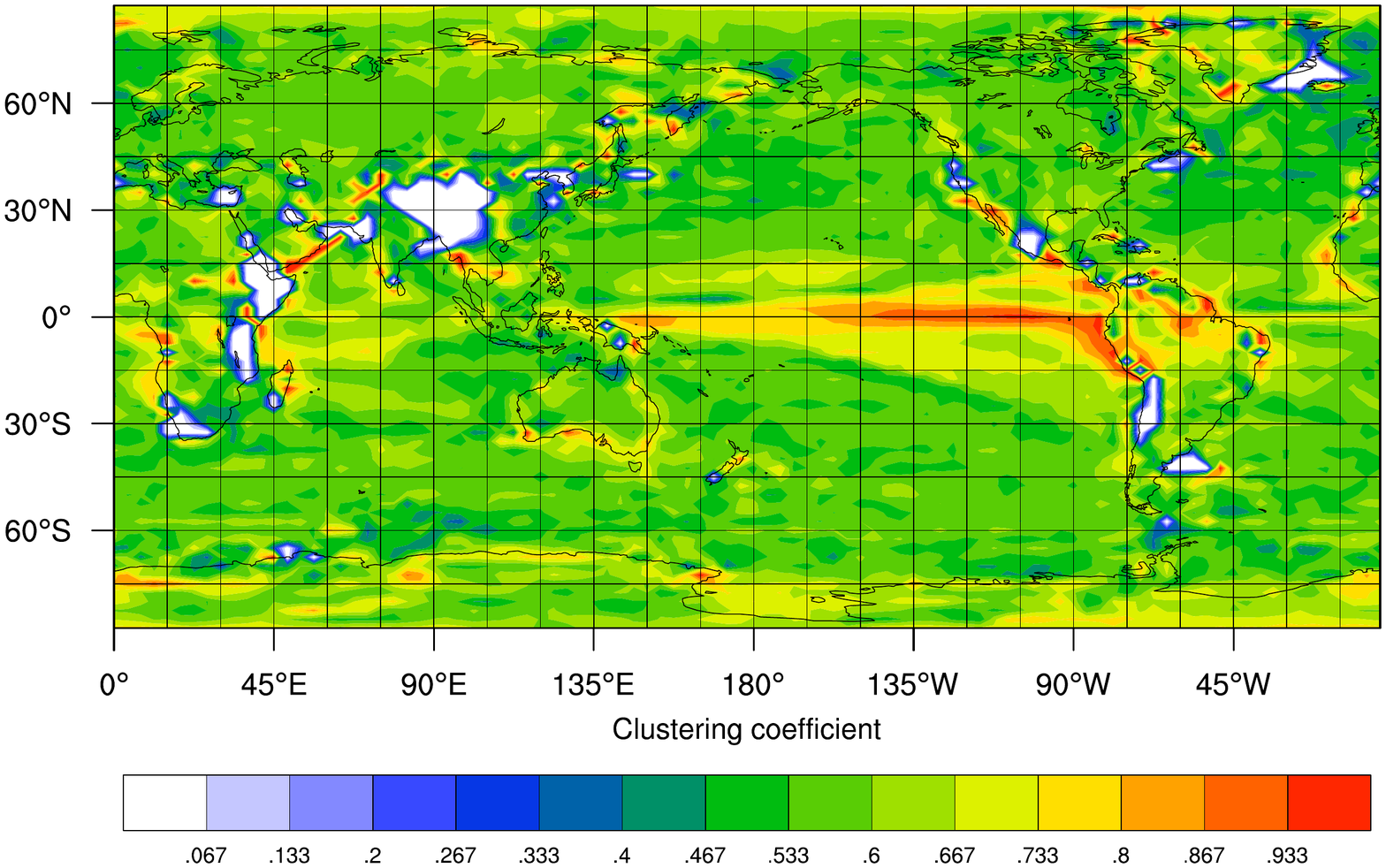}
\label{fig:ComparisonCSub1}
}
\subfigure[]{
\includegraphics[width=\textwidth, viewport=35 85 715 510, clip]{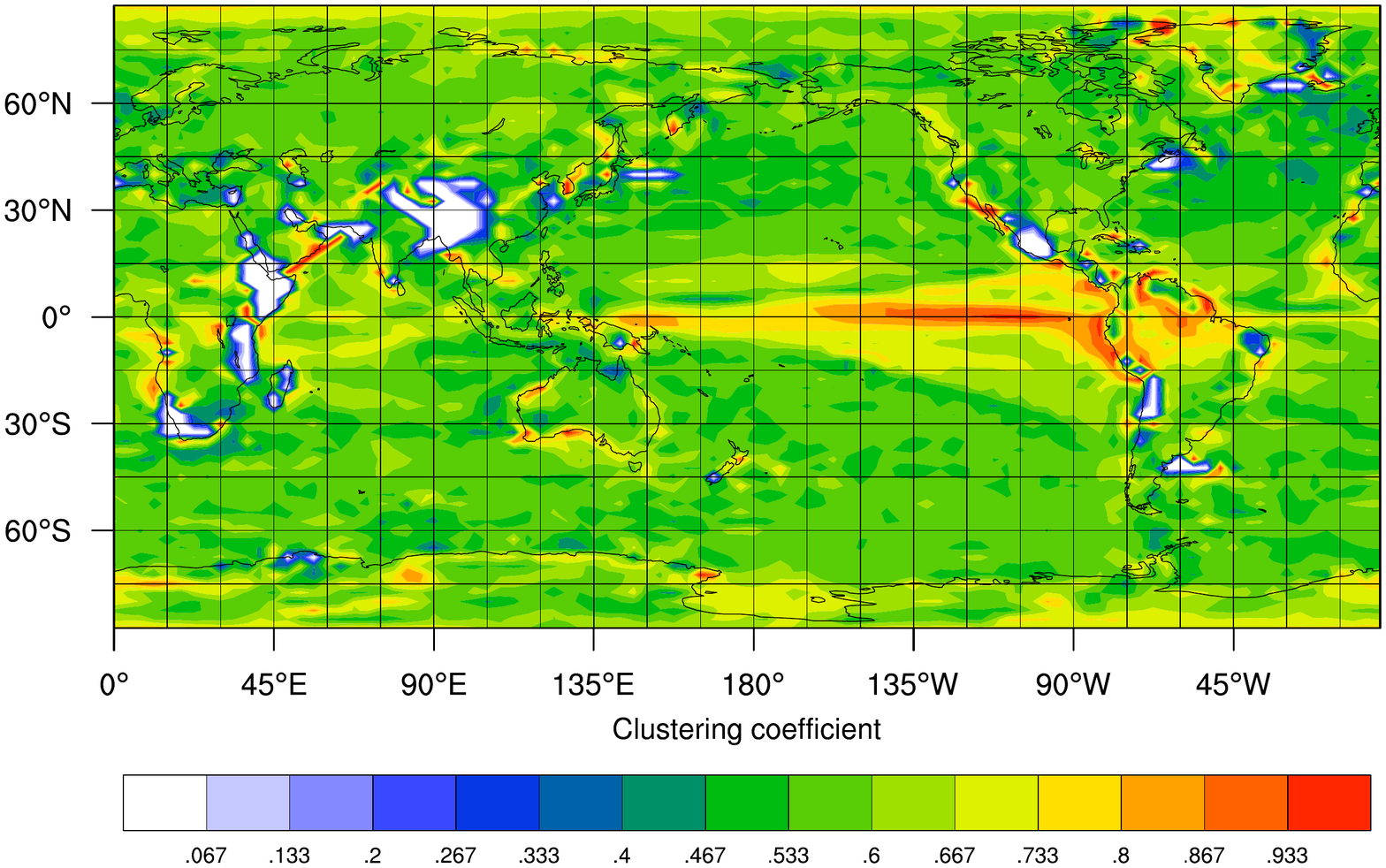}
\label{fig:ComparisonCSub2}
}

\caption{\label{fig:ComparisonC}Local Watts-Strogatz clustering coefficient fields for global HadCM3 SAT networks at $\rho = 0.005$ (linear color scale) obtained using a) Pearson correlation, b) mutual information. The rank order correlation between the two fields is $r_s^{\mathcal{C}}(0.005) = 0.81$.}
\end{figure}

The local and global clustering coefficients also reveal a high degree of similarity on the mesoscopic topological scale. Analogously to $AWC$, the local clustering coefficient fields are nearly indistinguishable (Fig. \ref{fig:ComparisonC}). However, the largest deviations appear to cluster along coastlines (Fig. \ref{fig:DifferenceFields02}). This interesting finding can be understood by considering the qualitatively different dynamics of SAT over oceans and continents, e.g. the on average much larger seasonal variability over continents. Along coastlines, the correlation length of the SAT field is thus smaller than that expected over continents or the ocean away from the coast. Hence Pearson correlation and mutual information have a higher probability to disagree on the existence of edges between spatially adjacent vertices (local edges) along the coastline. These local and mesoscopic deviations in network structure are detected by the local correlation coefficient $C_v$, that is by design particularly sensitive on the mesoscopic topological scale (Sect. \ref{Climatological_Interpretation}).

The rank order correlation coefficient reaches a maximum between $\rho=0.005$ and $\rho=0.01$ and decays for larger edge densities (Fig. \ref{fig:HadCM3_SAT_Comp04}). We obtain high values for $\rho=0.005$ and $\rho=0.01$ (Table \ref{ComparisonTable}). The global clustering coefficients show only small deviations of $\mathcal{O}(10^{-2})$ at all edge densities considered (Fig. \ref{fig:HadCM3_SAT_Comp02}). We get $\mathcal{C}^P(0.005)=0.682$, $\mathcal{C}^M(0.005)=0.678$ and $\mathcal{C}^P(0.01)=0.657$, $\mathcal{C}^M(0.01)=0.668$. The local clustering coefficient field is close to normally distributed.

\subsection{Global comparison \label{Global_comparison}}

\begin{figure}[tp]
\centering

\subfigure[]{
\includegraphics[width=\textwidth, viewport=35 85 715 510, clip]{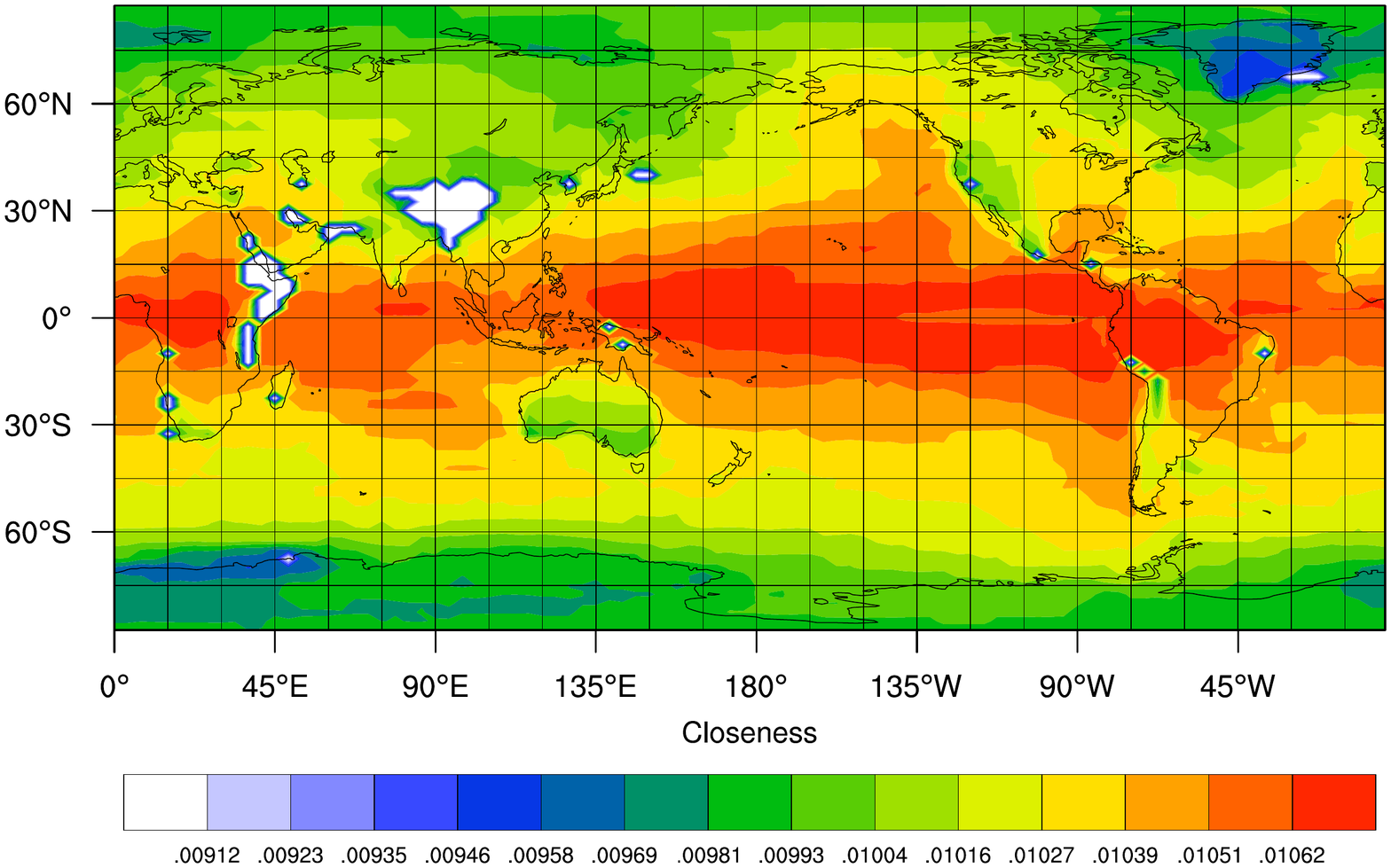}
\label{fig:ComparisonCCSub1}
}
\subfigure[]{
\includegraphics[width=\textwidth, viewport=35 85 715 510, clip]{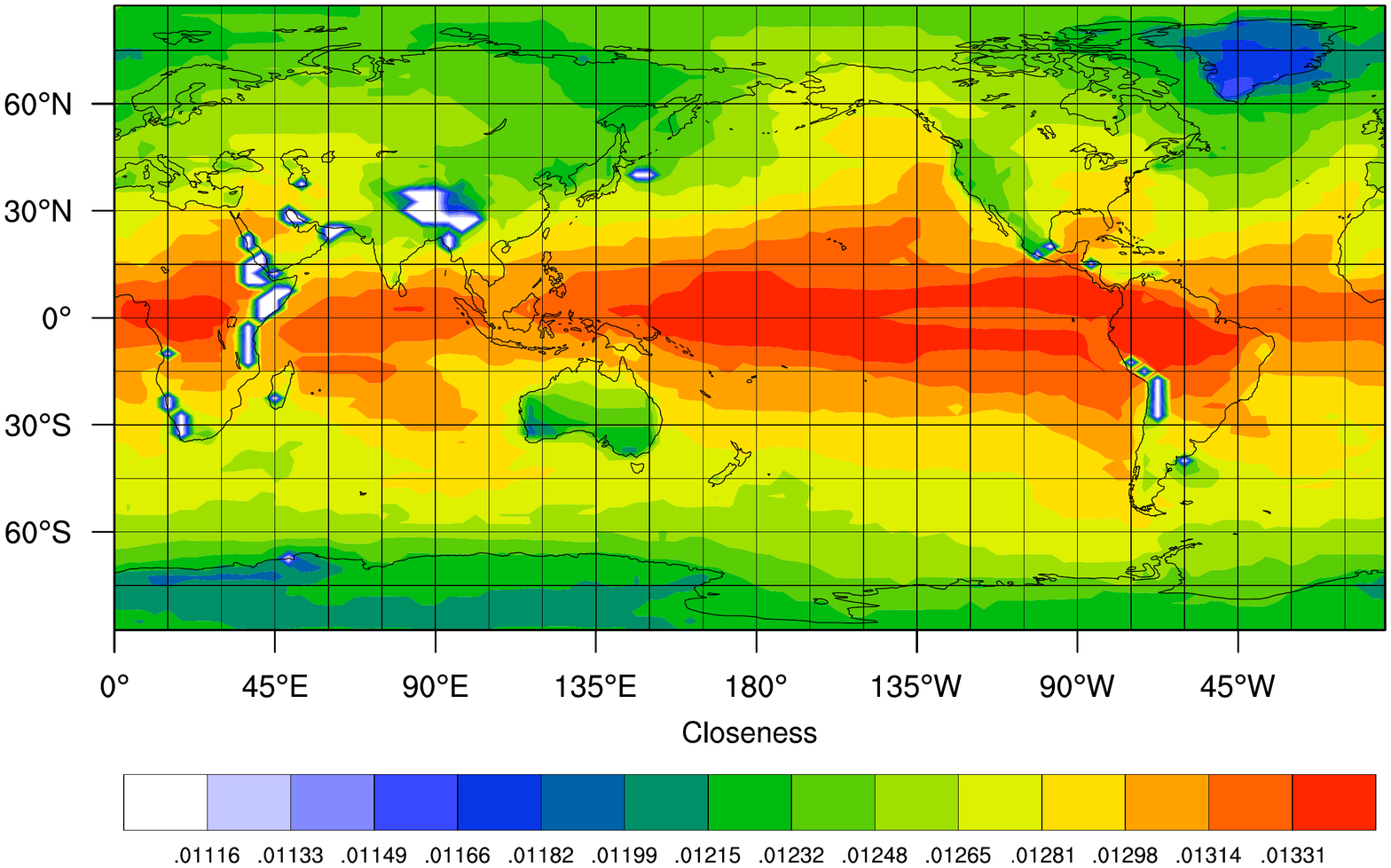}
\label{fig:ComparisonCCSub2}
}

\caption{\label{fig:ComparisonCC}Closeness centrality field for global HadCM3 SAT networks at $\rho = 0.005$ (linear color scale) obtained using a) Pearson correlation, b) mutual information. The rank order correlation between the two fields is $r_s^{CC}(0.005) = 0.98$. The white regions on the map correspond to vertices that are disconnected from the network's giant component.}
\end{figure}

\begin{figure}[tp]
\centering

\subfigure[]{
\includegraphics[width=\textwidth, viewport=35 85 715 510, clip]{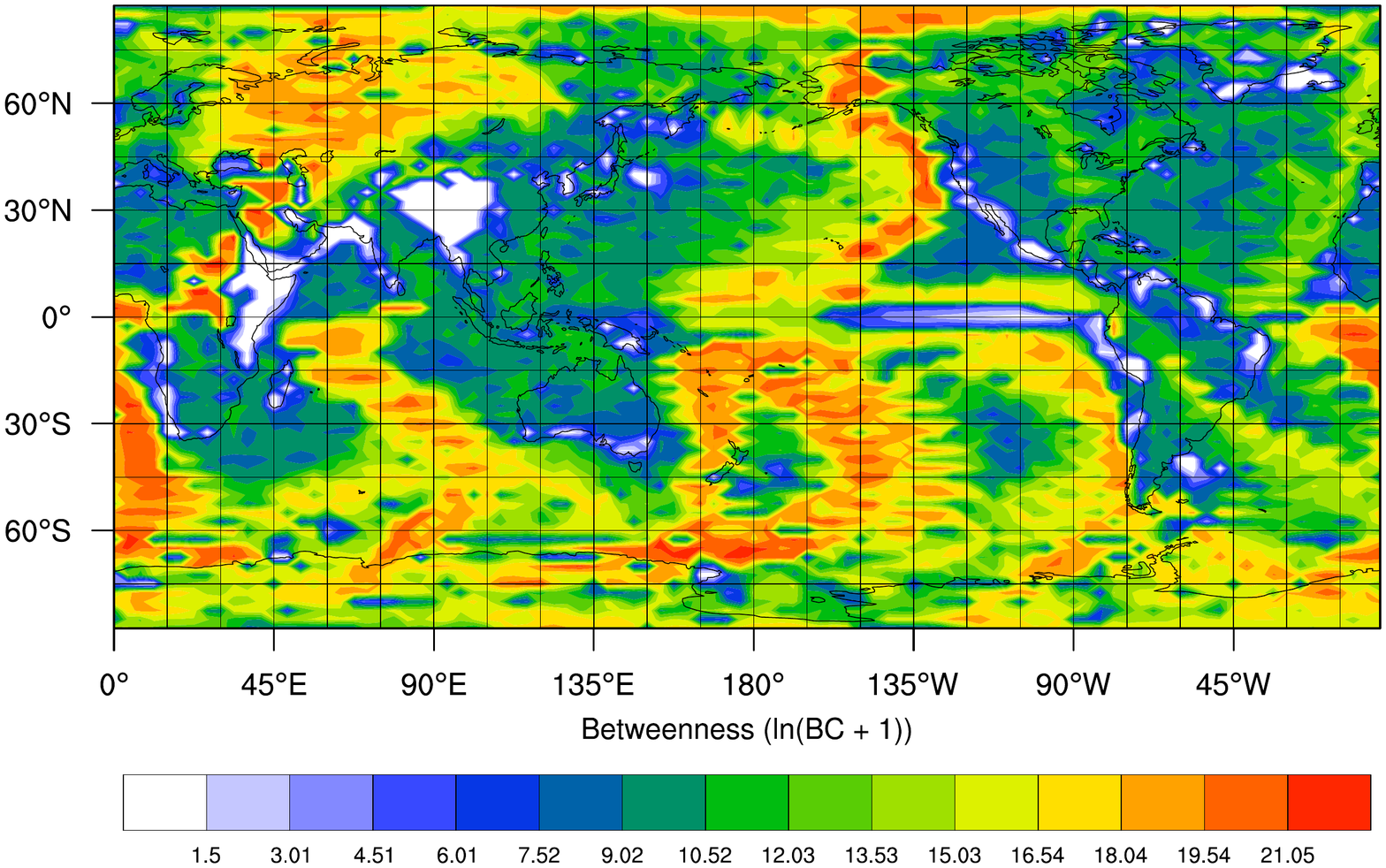}
\label{fig:ComparisonBCSub1}
}
\subfigure[]{
\includegraphics[width=\textwidth, viewport=35 85 715 510, clip]{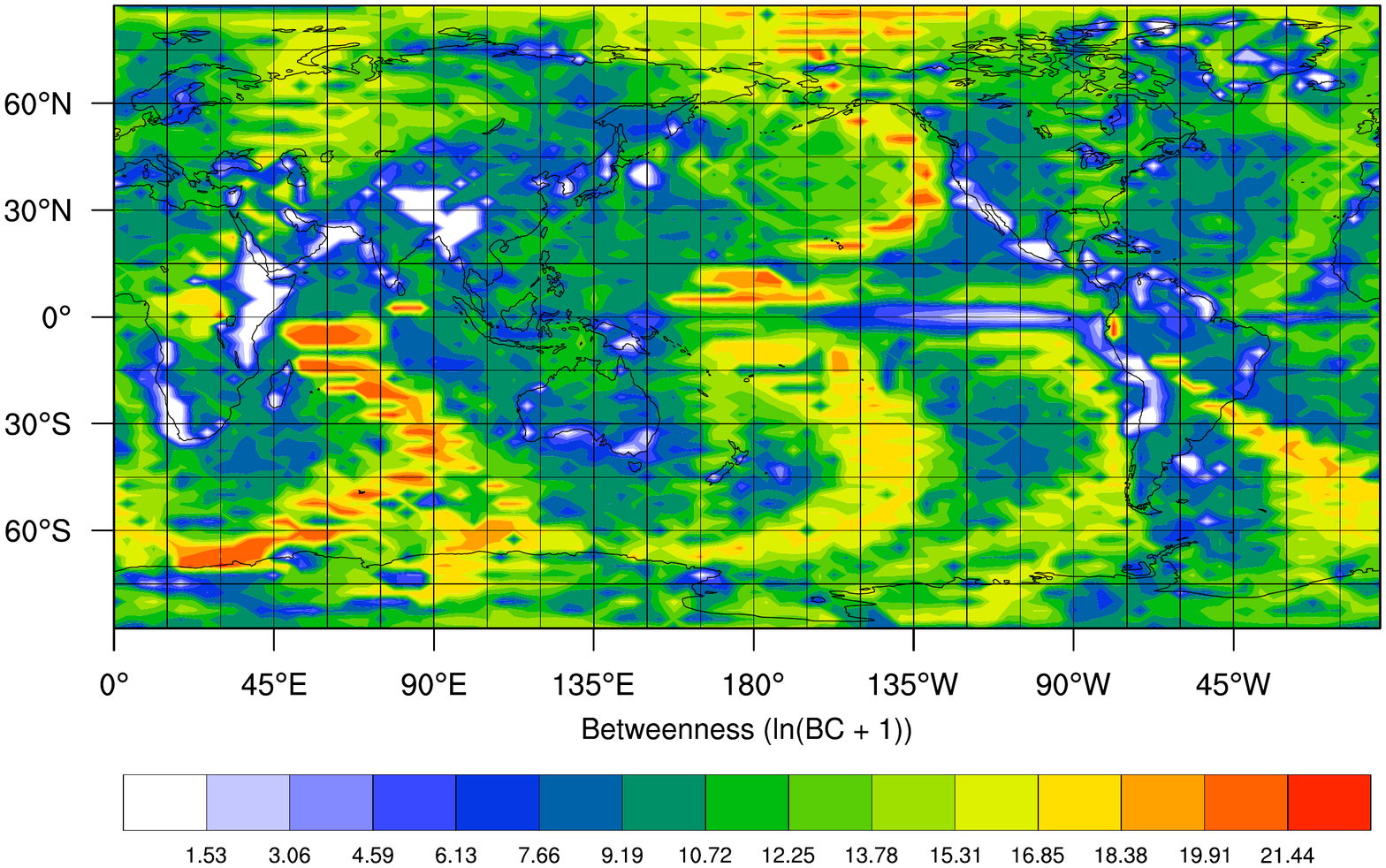}
\label{fig:ComparisonBCSub2}
}

\caption{\label{fig:ComparisonBC}Betweenness centrality fields for global HadCM3 SAT networks at $\rho = 0.005$ (logarithmic color scale) obtained using a) Pearson correlation, b) mutual information. The rank order correlation between the two fields is $r_s^{BC}(0.005) = 0.70$.}
\end{figure}

\begin{figure}
\centering

\subfigure[]{
\includegraphics[width= 0.48\textwidth, viewport=35 78 715 510, clip]{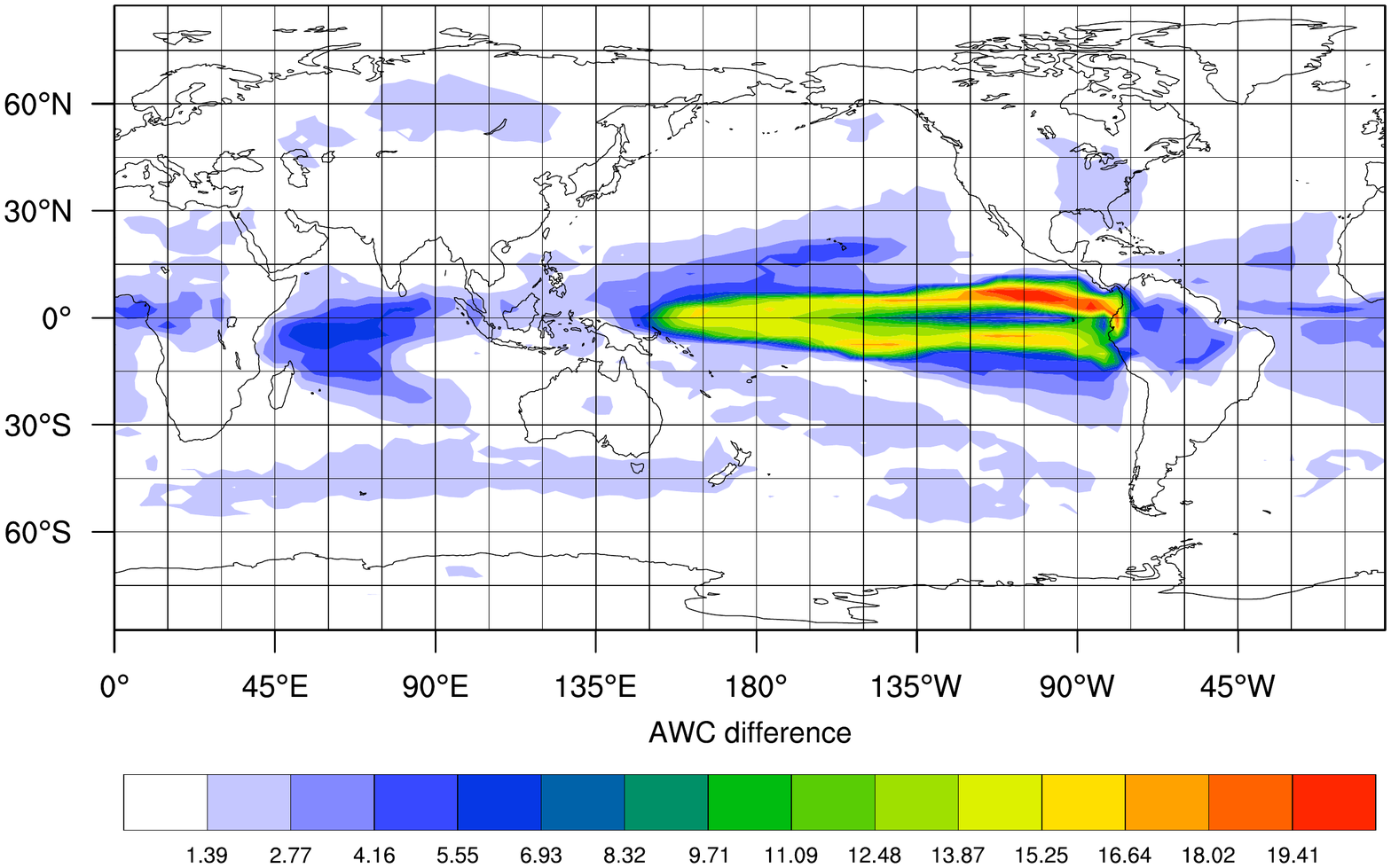}
\label{fig:DifferenceFields01}
}
\subfigure[]{
\includegraphics[width= 0.48\textwidth, viewport=35 78 715 510, clip]{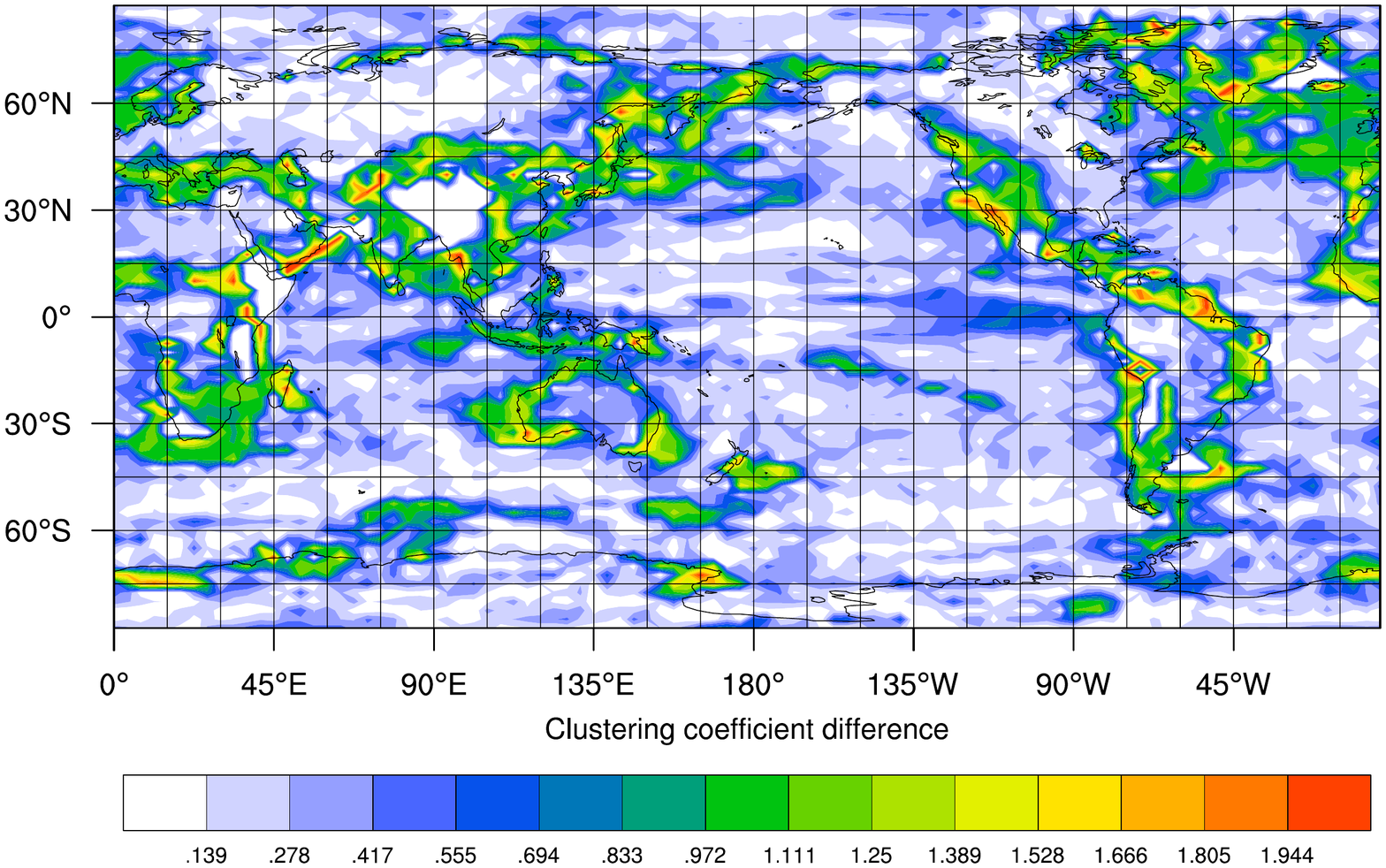}
\label{fig:DifferenceFields02}
}
\subfigure[]{
\includegraphics[width= 0.48\textwidth, viewport=35 78 715 510, clip]{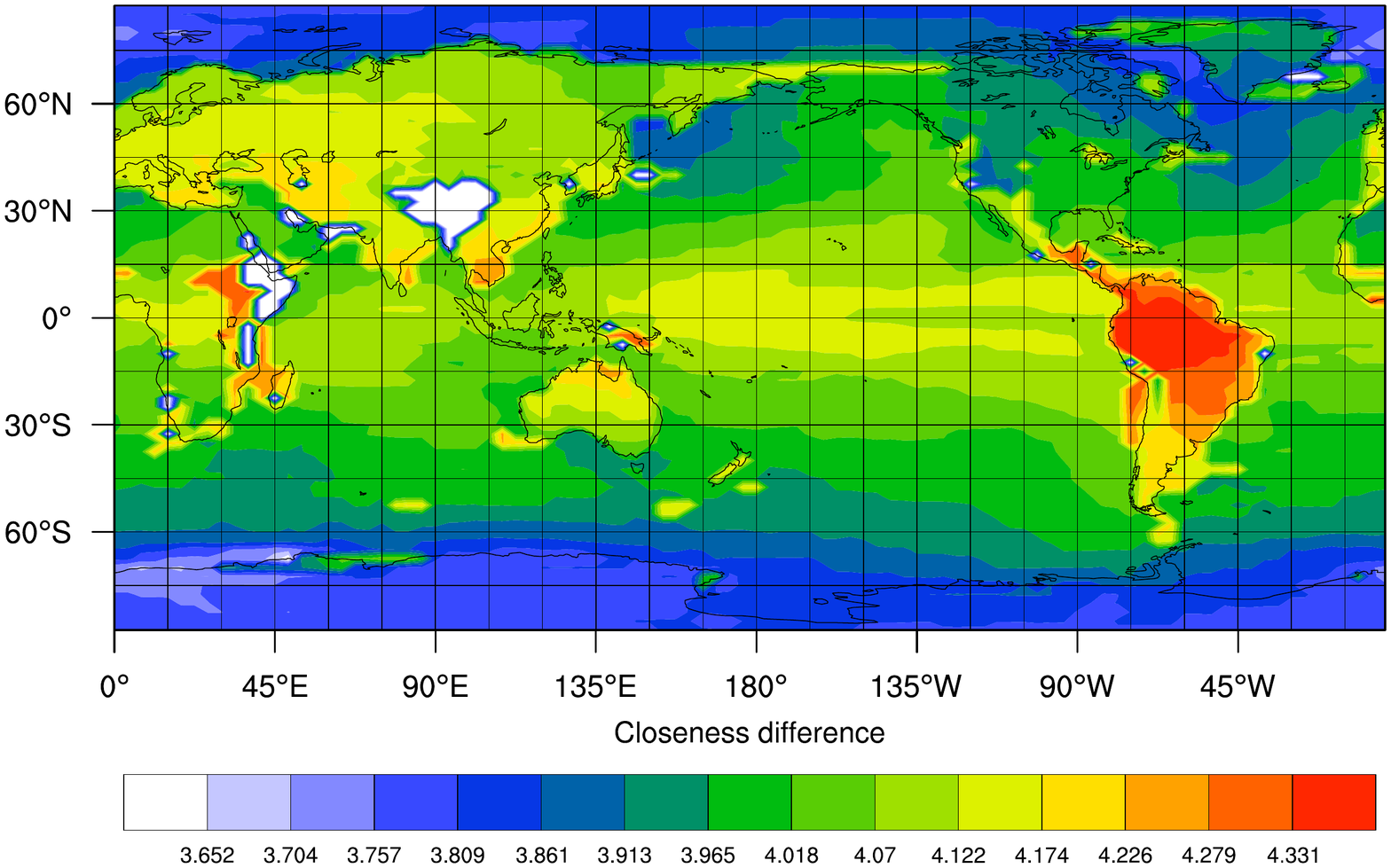}
\label{fig:DifferenceFields03}
}
\subfigure[]{
\includegraphics[width= 0.48\textwidth, viewport=35 78 715 510, clip]{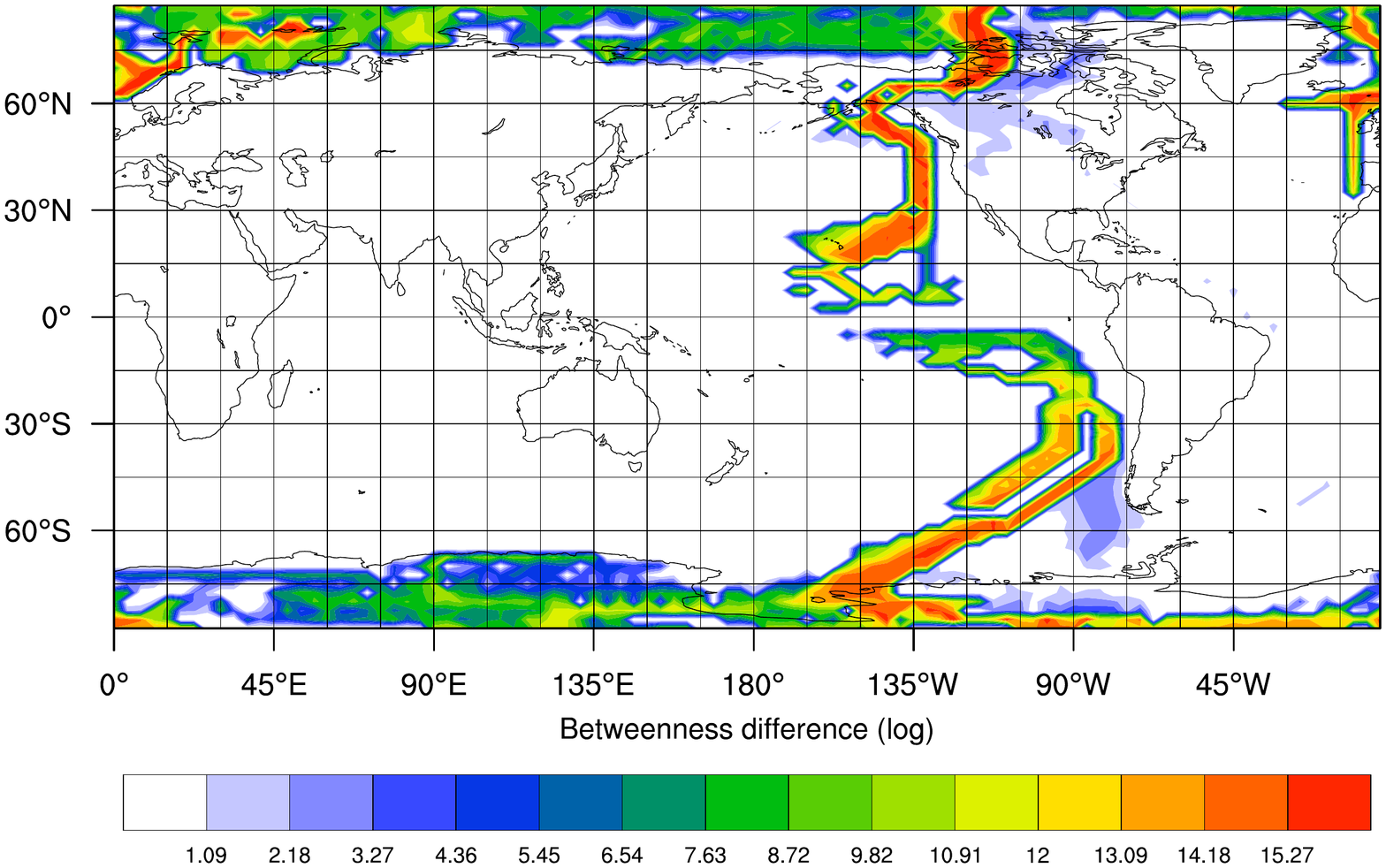}
\label{fig:DifferenceFields04}
}

\caption{\label{fig:DifferenceFields}Normalized difference fields $\Delta g_v = |g^P_v - g^M_v| / \sqrt{\left<g^P_w\right>_w \left<g^M_w\right>_w}$ of network measure fields $g^P_v$ and $g^M_v$, calculated from Pearson correlation and mutual information HadCM3 SAT climate networks. (a) Area weighted connectivity, (b) local clustering coefficient, (c) closeness and (d) betweenness.}
\end{figure}

\begin{figure}
\centering

\subfigure[]{
\includegraphics[width= 0.48\textwidth]{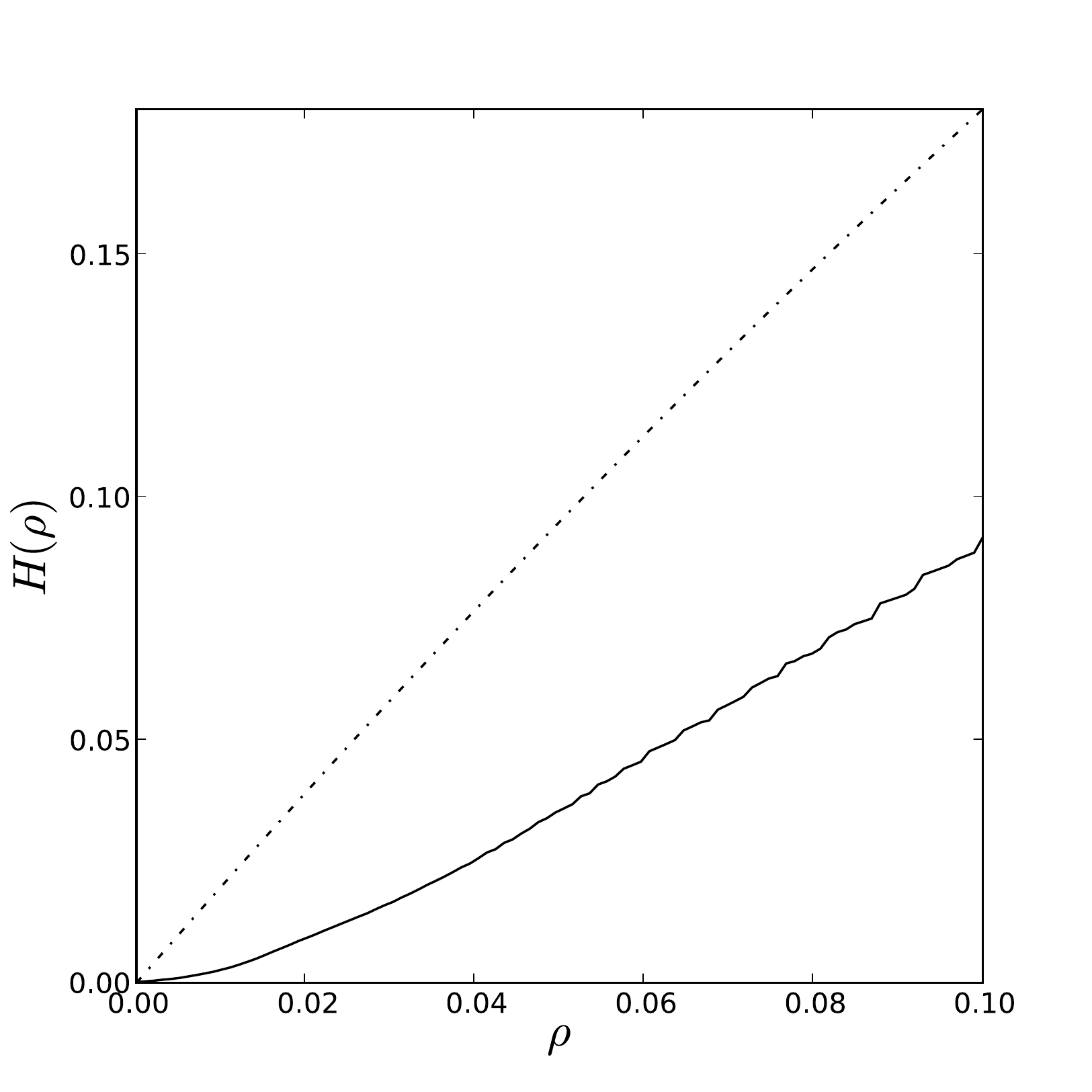}
\label{fig:HadCM3_SAT_Comp01}
}
\subfigure[]{
\includegraphics[width= 0.48\textwidth]{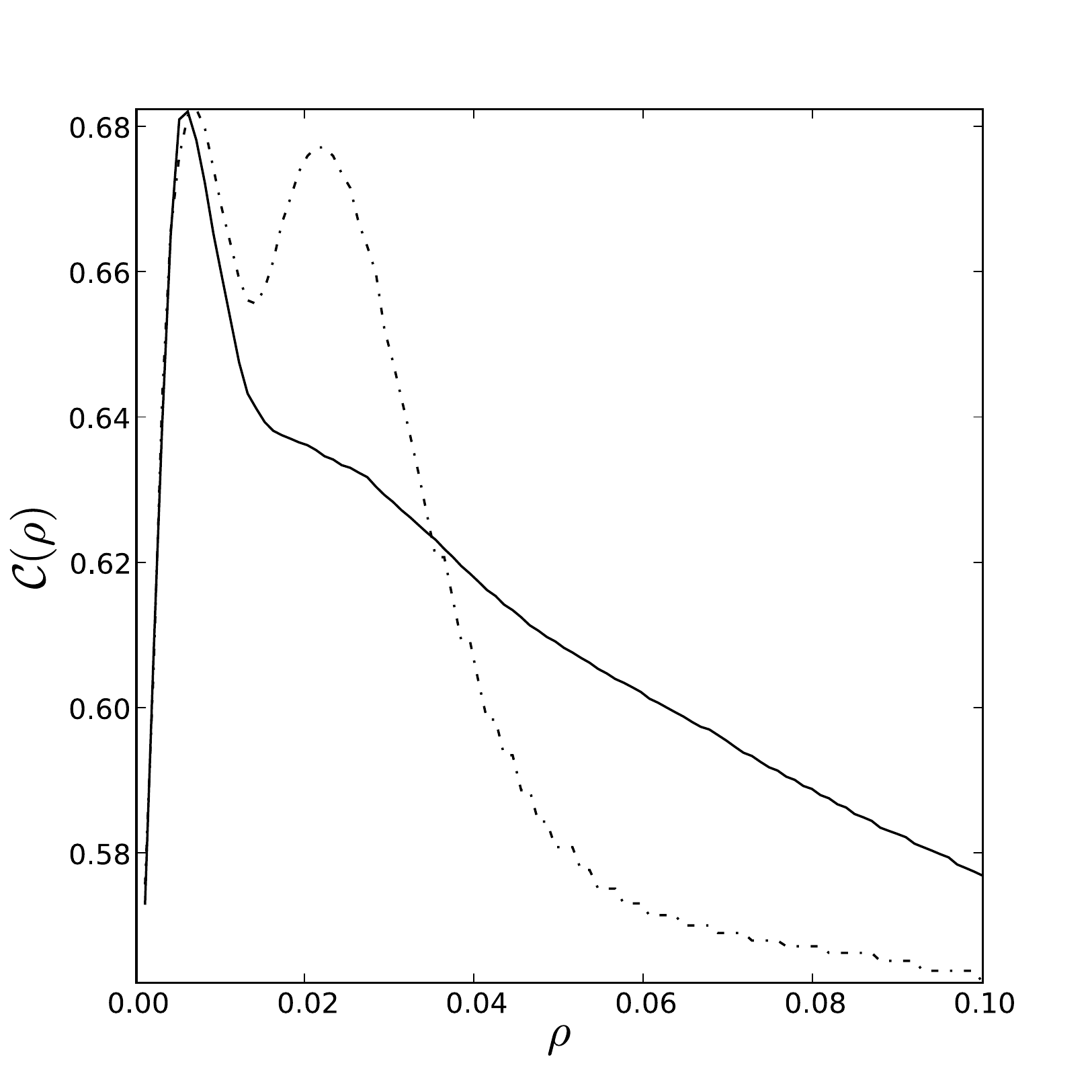}
\label{fig:HadCM3_SAT_Comp02}
}
\subfigure[]{
\includegraphics[width= 0.48\textwidth]{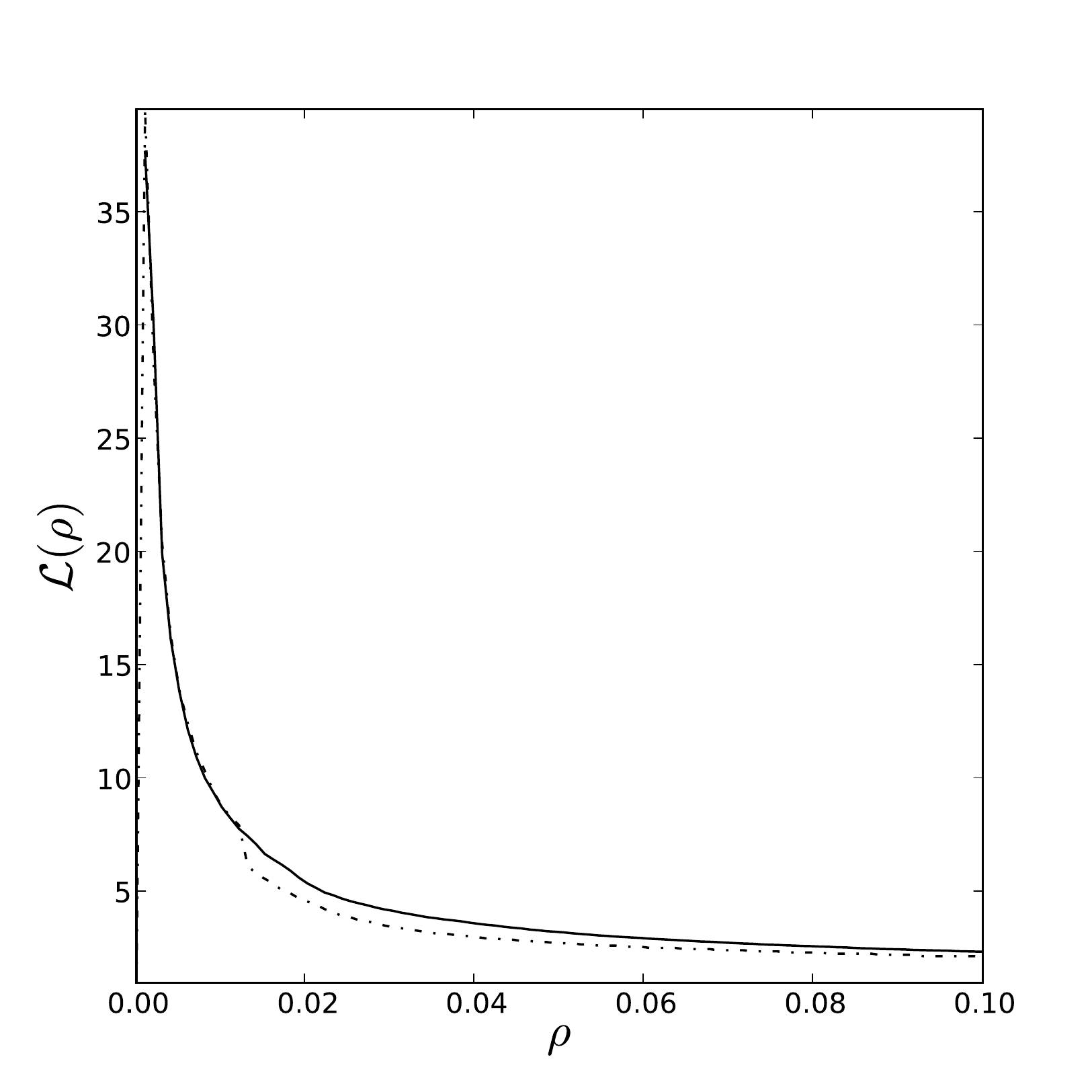}
\label{fig:HadCM3_SAT_Comp03}
}
\subfigure[]{
\includegraphics[width= 0.48\textwidth]{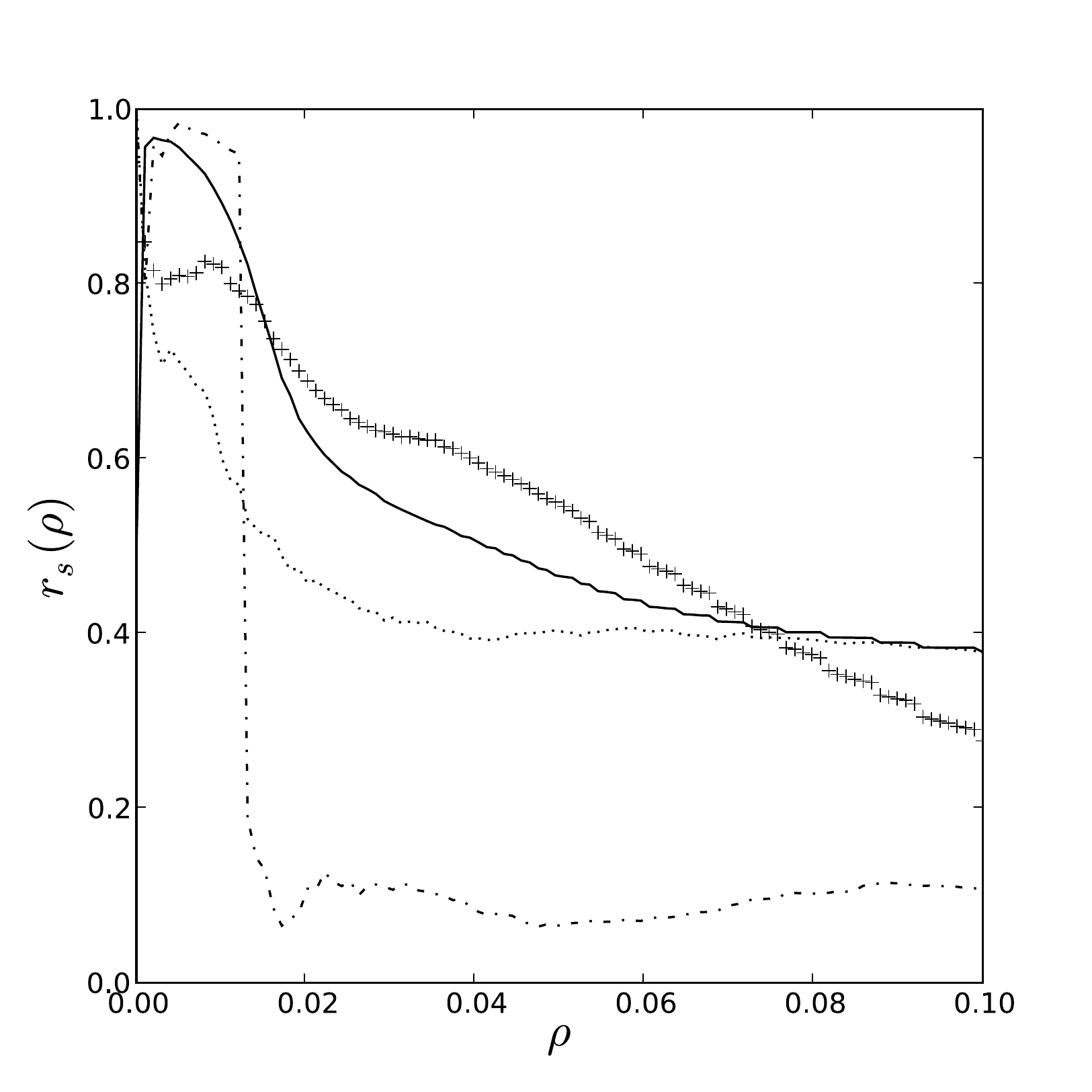}
\label{fig:HadCM3_SAT_Comp04}
}

\caption{\label{fig:HadCM3_SAT_Comp}Results for the quantitative comparison of Pearson correlation and mutual information climate networks for the global HadCM3 SAT data set, shown as a function of edge density $\rho$ (100 edge density steps). (a) Hamming distance $H(\rho)$ (continuous line) and expected random Hamming distance $H^R(\rho)$ (dashed line) between the two networks. The expected deviations from $H^R(\rho)$ are of $\mathcal{O}(10^{-4})$ (Sect. \ref{Local_comparison}). (b) Global clustering coefficient $\mathcal{C}^P(\rho)$ of the Pearson correlation (continuous line) and $\mathcal{C}^M(\rho)$ of the mutual information network (dashed line). (c) Average path length $\mathcal{L}^P(\rho)$ of the Pearson correlation (continuous line) and $\mathcal{L}^M(\rho)$ of the mutual information network (dashed line). (d) Spearman rank order correlation coefficients $r_s^{AWC}(\rho)$ for the area weighted connectivity (continuous line), $r_s^{\mathcal{C}}(\rho)$ for the local clustering coefficient (crosses), $r_s^{CC}(\rho)$ for the closeness centrality (dash-dotted line) and $r_s^{BC}(\rho)$ for the betweenness centrality fields (dotted line).}
\end{figure}

We observe more interesting behavior at the global topological scale. Closeness centrality at  $\rho=0.005$ does not deviate much qualitatively and quantitatively across the two types of networks considered (Fig. \ref{fig:ComparisonCC}), the largest differences are detected in the tropics with a tendency to decrease with latitude towards the poles, and most notably over South America (Fig. \ref{fig:DifferenceFields03}). The betweenness centrality field shows more pronounced qualitative regional differences (Fig. \ref{fig:ComparisonBC}). For example, note the differing high betweenness structures over the oceans, particularly over the East Pacific, the North Atlantic and arctic regions (Fig. \ref{fig:DifferenceFields04}). The rank order correlation coefficients $r_s^{CC}$ and $r_s^{BC}$ decay more quickly than the ones on the local and mesoscopic topological scale and fluctuate around values of $r_s^{CC}\approx 0.1$ and $r_s^{BC} \approx 0.4$ for larger edge densities (Fig. \ref{fig:HadCM3_SAT_Comp04}). At $\rho=0.005$ and $\rho=0.01$, $r_s^{BC}$ is notably smaller than the Spearman's Rho of the other fields considered, while $r_s^{CC}$ is close to unity (Table \ref{ComparisonTable}). Confirming earlier studies, we find that betweenness follows a fat tailed PDF \cite{goh2002csf}, whereas the closeness field is normally distributed.

These results indicate, that betweenness centrality may quantify the local differences between networks constructed using Pearson correlation and mutual information at the global topological scale, that could be traces of nonlinear physical processes in the climate system. That the greatest deviations are found between the betweenness centrality fields is plausible, because betweenness is by definition a very sensitive measure and can locally depend heavily on the existence or non-existence of a small number of edges in the network \cite{albert2004svn}. Consider for example a small set of edges, that are the only connections between two large communities in a network. The vertices on either end of these edges have a high betweenness centrality, because all shortest paths between the two communities must contain them. If the bridging edges are removed, the betweenness centrality of the beachhead vertices must decrease significantly, since they can now only participate in shortest paths within their own community. This sensitivity of betweenness leads to a large dynamic range of $20$ orders of magnitude for the global HadCM3 SAT network, that calls for a logarithmic scale to properly visualize the betweenness distribution (Fig. \ref{fig:ComparisonBC}).

The average path length (Fig. \ref{fig:HadCM3_SAT_Comp03}) agrees closely, with deviations of $\mathcal{O}(10^{-1})$. We obtain $\mathcal{L}^P(0.005)=13.4$, $\mathcal{L}^M(0.005)=13.5$ and $\mathcal{L}^P(0.01)=8.5$, $\mathcal{L}^M(0.01)=8.5$.

\begin{figure}
\centering

\subfigure[]{
\includegraphics[width= 0.48\textwidth]{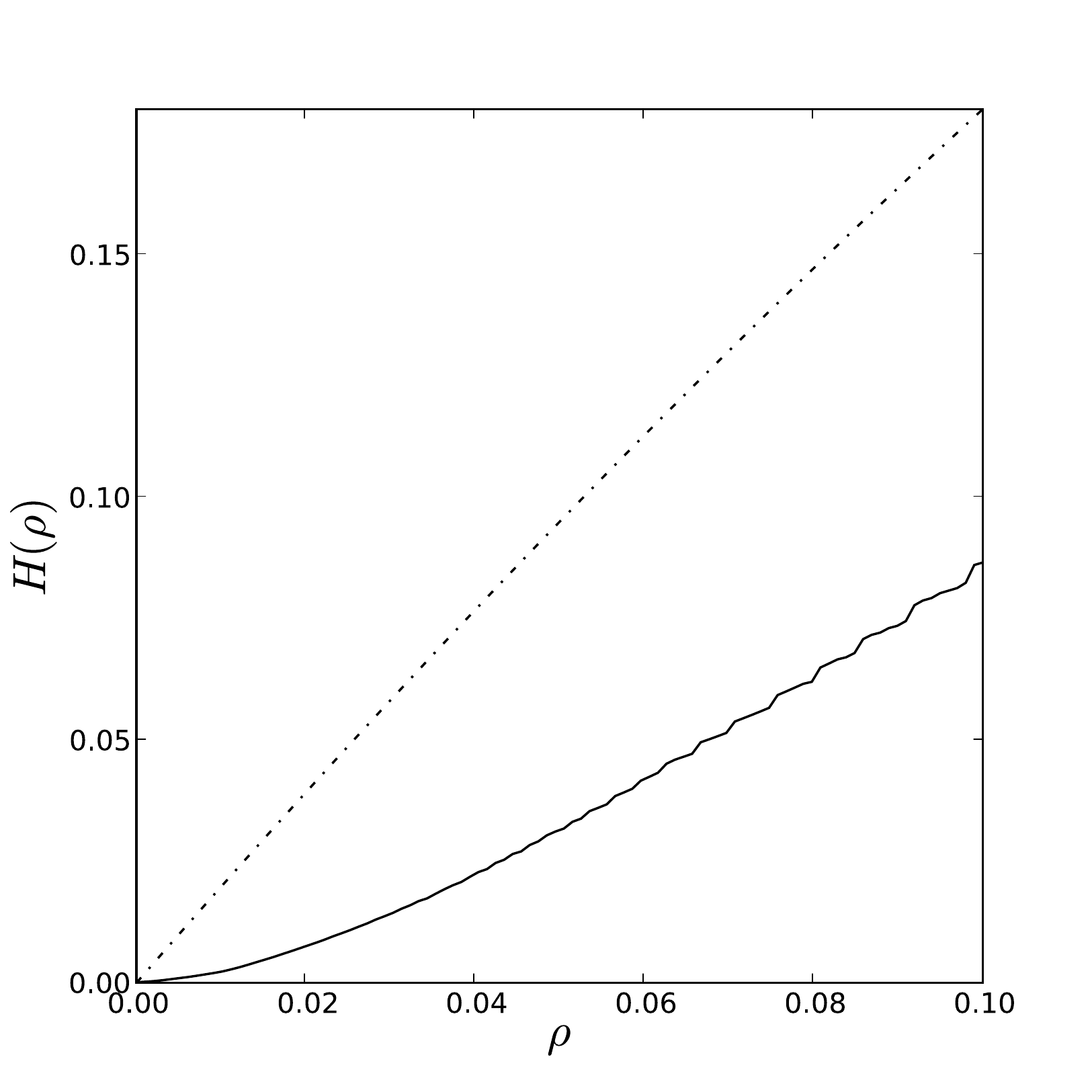}
\label{fig:Reanalysis_SAT_Comp01}
}
\subfigure[]{
\includegraphics[width= 0.48\textwidth]{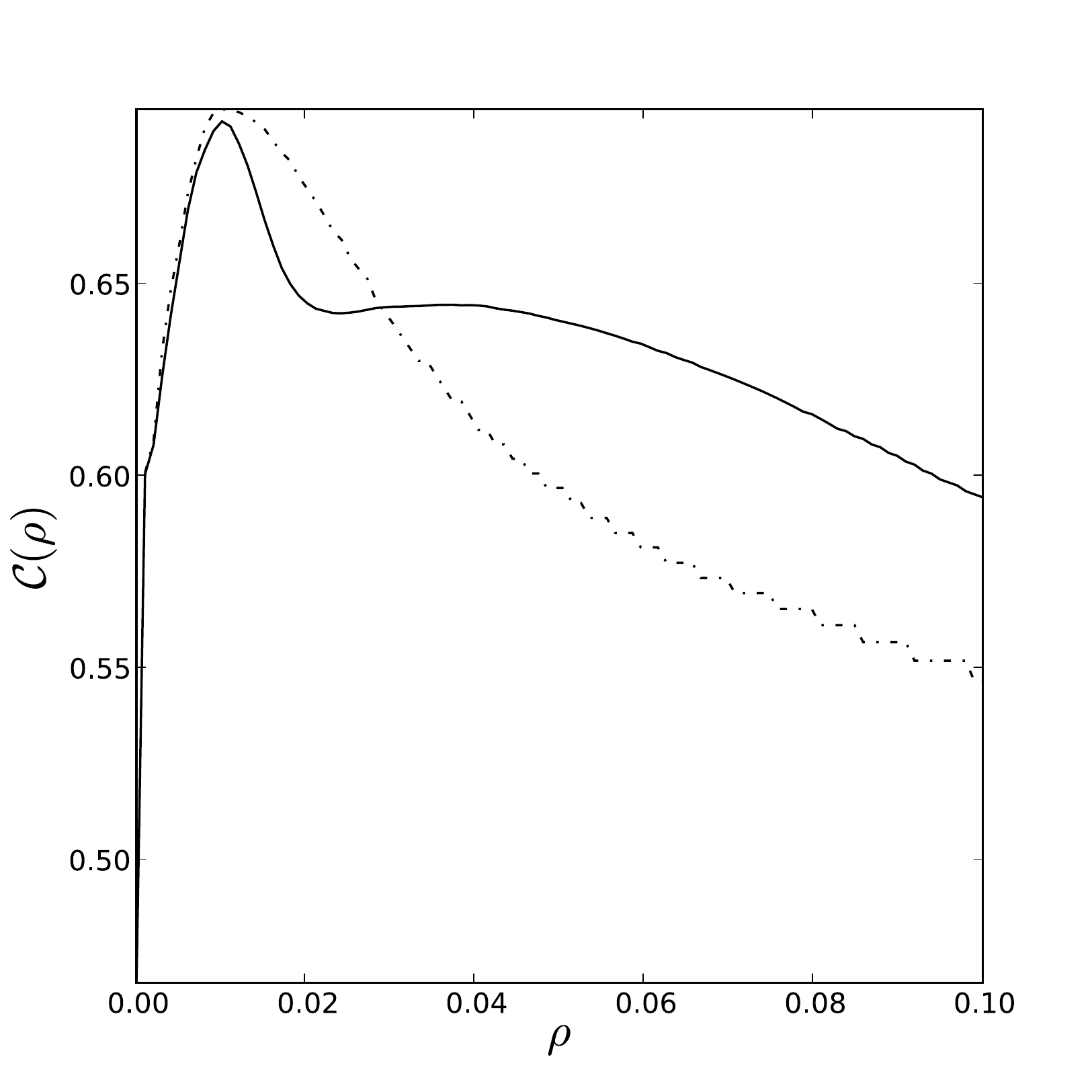}
\label{fig:Reanalysis_SAT_Comp02}
}
\subfigure[]{
\includegraphics[width= 0.48\textwidth]{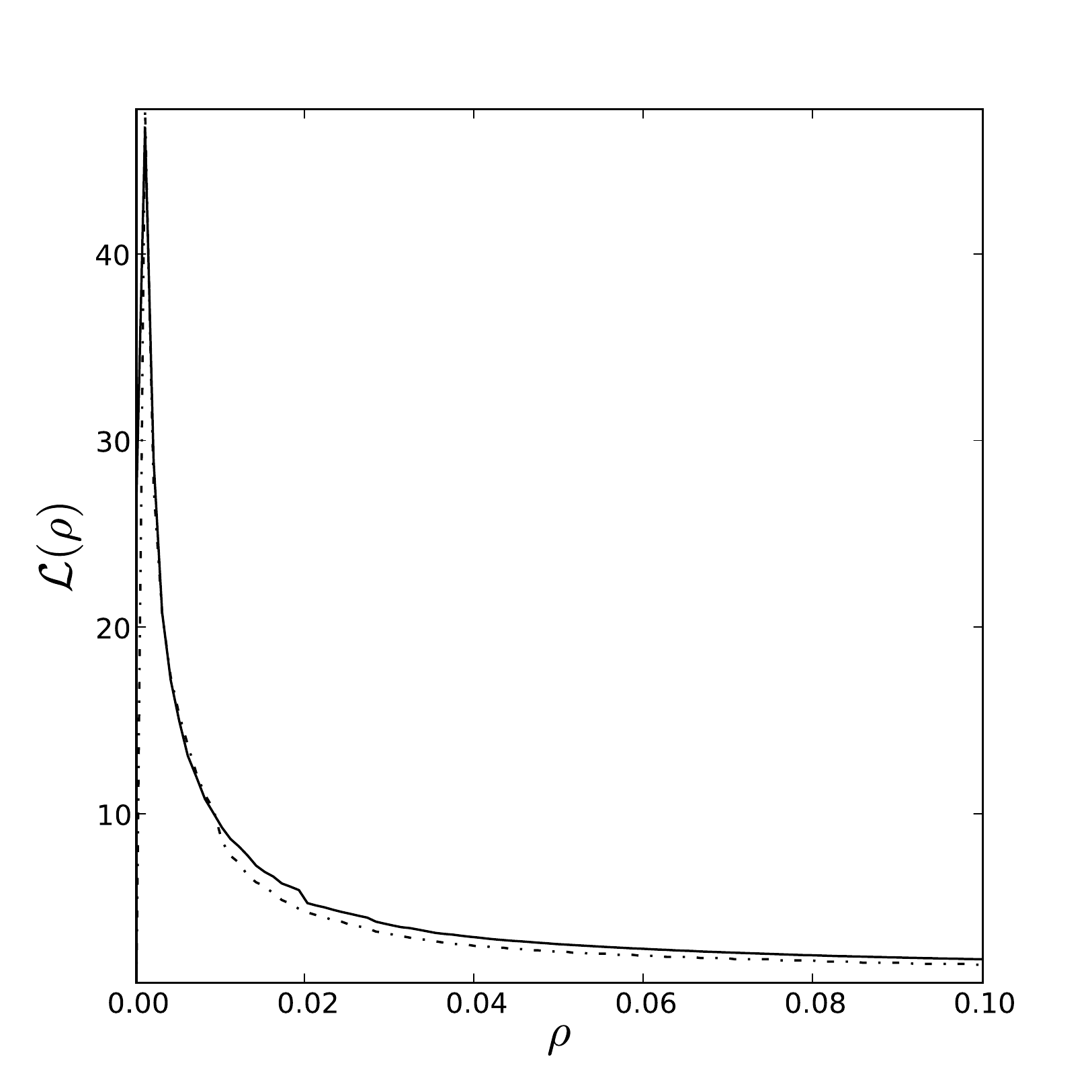}
\label{fig:Reanalysis_SAT_Comp03}
}
\subfigure[]{
\includegraphics[width= 0.48\textwidth]{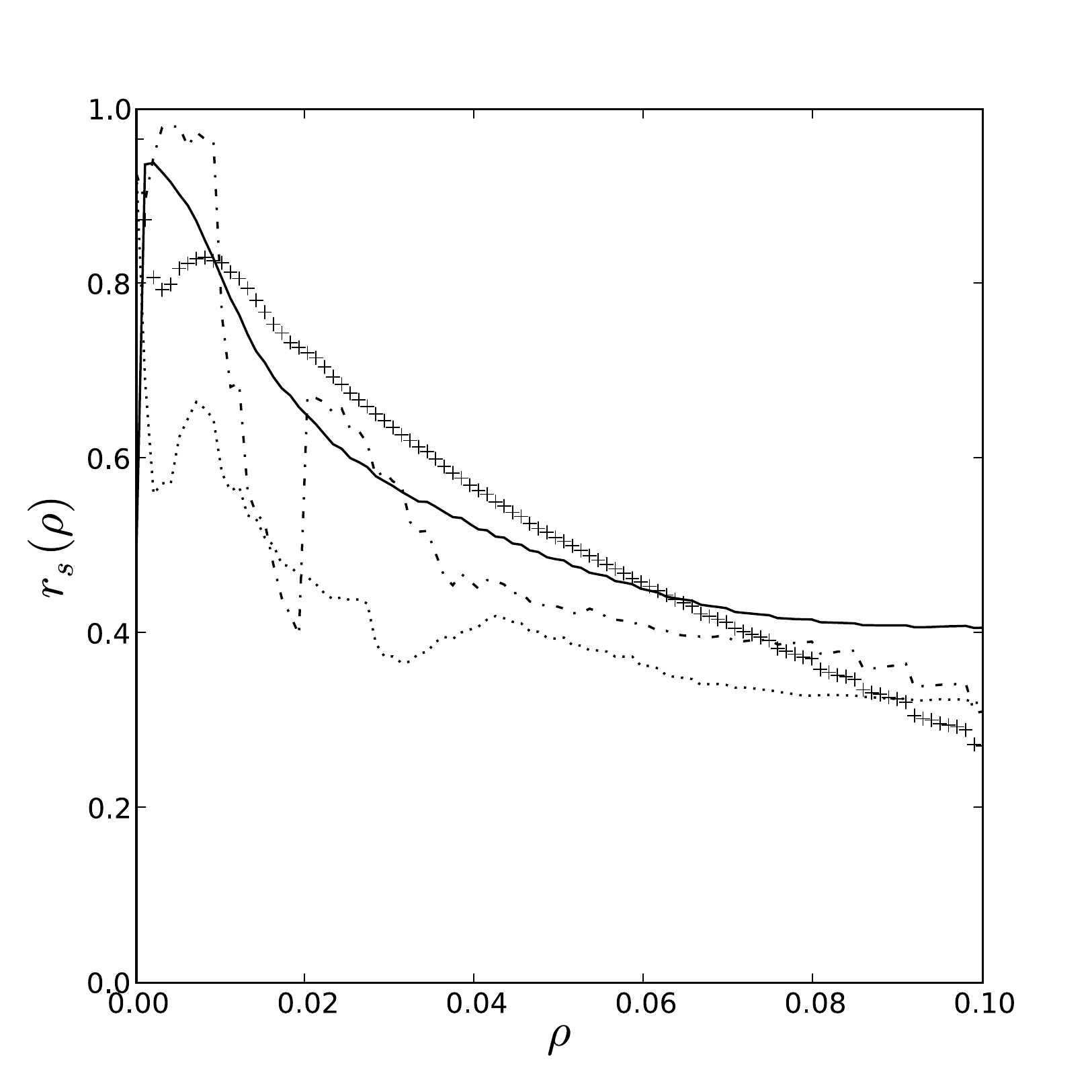}
\label{fig:Reanalysis_SAT_Comp04}
}

\caption{\label{fig:Reanalysis_SAT_Comp}This figure shows the same statistics as Fig. \ref{fig:HadCM3_SAT_Comp}, but evaluated for the global NCEP/NCAR reanalysis SAT data set.}
\end{figure}

\subsection{Climatological interpretation} \label{Climatological_Interpretation}

We give brief climatological interpretations of the network properties unveiled by our approach, since the main aim of this study is the comparison of linear and nonlinear climate network construction methods (Sect. \ref{Construction}). Super-nodes found in the AWC field (Fig. \ref{fig:ComparisonAWC}) over the tropics and locally the mid-latitudes, were shown to be related to major atmospheric teleconnection patterns \cite{tsonis2008jclim}. For example, the region of increased AWC in the North East Pacific is associated to the well-known Pacific North-American (PNA) pattern \cite{wallace1981tgh}. The El Ni\~no cold tongue in the tropical East Pacific is clearly visible in the AWC field, as well as in all other fields considered (Fig. \ref{fig:ComparisonC}, \ref{fig:ComparisonCC} and \ref{fig:ComparisonBC}).

The local clustering coefficient is found to be of $\mathcal{O}(1)$ in a connected region in the equatorial Pacific as well as locally along continental coastlines (Fig. \ref{fig:ComparisonC}). The former indicates a high degree of dynamical similarity in the tropical Pacific \cite{tsonis2006,tsonis2008jclim}, that is possibly related to ENSO. The latter are more likely to be a signature of our climate network construction method along the coastline and visible on the mesoscopic scale only, that we discuss in Sect. \ref{Mesoscopic_comparison}.

The contouring of the closeness field (Fig. \ref{fig:ComparisonCC}) nicely shows the latitudinally growing influence of the Coriolis force. Pressure gradient forces are balanced by the Coriolis force in the mid-latitudes for large scale atmospheric flows. This balance vanishes in the tropics, because the Coriolis force decays as $\sin(\lambda)$ when latitude $\lambda$ approaches the equator. The closeness field also shows that the tropics form the center of the SAT climate network, the associated vertices being topologically closer to the rest of the network than vertices in the mid-latitudes and arctic regions. This finding can be explained by considering the comparably regular dynamics of the tropical SAT field leading to many edges between tropical vertices and the more irregular dynamics in the mid-latitudes and arctic regions that results in fewer edges within the mid-latitudes and arctic as well as between these regions and the tropics \cite{vonbloh2005ltp}. In a global climate network, it is hence more probable to find shorter shortest paths starting from tropical vertices, while shortest paths originating in mid-latitude and arctic vertices are on average longer. Moreover, we point out the lower closeness over Australia and Greenland indicating that these land-masses also form pronounced clusters in the SAT climate network, even though the local clustering coefficient field shows that they are not as highly locally interconnected as the equatorial Pacific. These differences in local connectedness among the detected dynamical clusters are caused by the qualitatively different dynamics over land and oceans (Sect. \ref{Mesoscopic_comparison}). The land-sea difference is globally detected by closeness centrality and AWC: Vertices over land masses are found to be on average less well connected and topologically more remote than those over the oceans.

We observe highly localized linear structures in the betweenness field (Fig. \ref{fig:ComparisonBC}), some of which appear to resemble major surface ocean currents such as the California and Peru currents following the western coastline of the Americas, or the East Greenland, Norwegian and Canary currents. Note that some of these current resembling structures are particularly visible in the betweenness difference field (Sect. \ref{Global_comparison}), indicating that nonlinear processes might be involved in the formation of some of the structures. In analogy to the major communication channels of the internet, we refer to these betweenness structures as the backbone of the climate network, because a large fraction of the dynamical information exchanged via topologically shortest paths between all possible pairs of vertices $\{i,j\}$ must pass the high betweenness regions. This is particularly true for information transported by advective processes, where the assumption of information traveling on shortest paths can be substantiated by extremalization principles. In our recent work we report the discovery of the backbone and its possible role in stabilizing the climate system \cite{Donges2008}.

Note that the region very close to the equator in the tropical East Pacific has a comparatively low AWC, closeness and betweenness, but a high local clustering coefficient. This indicates that this region forms a internally densely connected cluster in a network sense, i.e. it is dynamically highly interrelated but nearly detached from the rest of the network. We interpret it as a pronounced manifestation of the equatorial Coriolis barrier \cite{vallis2006aao}, that can also be observed weakly over the equatorial Indian and Atlantic Oceans.

In agreement with \cite{tsonis2004acn,tsonis2006,tsonis2008jclim} we find that Pearson correlation and mutual information climate networks possess properties of 'small-world' networks \cite{Watts1998,milgram1967swp}, i.e. a small average path length $\mathcal{L} \ll N$ and a large clustering coefficient of $\mathcal{O}(1)$ (Table \ref{ComparisonTable}, Fig. \ref{fig:HadCM3_SAT_Comp} and \ref{fig:Reanalysis_SAT_Comp}). Complex 'small-world' networks with comparable global properties are frequently found in nature, e.g. the internet, power grids, social and neural networks, and constitute the subject of study of an equally diverse collection of sciences. The small average path length can be explained by the influence of teleconnections. This indicates that perturbations of the regional dynamics (vertex dynamics) can on average quickly affect the whole globe via paths consisting of statistically highly interrelated pairs of regions (edges). It has been argued that this serves to stabilize the climate system and to enhance the information transfer within it \cite{tsonis2004acn,tsonis2006,tsonis2008jclim,tsonis2008tap}. If the climate network was only locally connected, in other words if all teleconnections were removed from it, the average path length would be of $\mathcal{O}(N)$ as that of a regular grid. The high clustering coefficient is due to the spatial continuity of the underlying physical fields (e.g. SAT), that leads to a prevalence of local triangles \cite{tsonis2008ecc}.


\section{Conclusions\label{Conclusions}}

In summary, we have performed a systematic study of the similarity of climate networks constructed using the linear Pearson correlation and the nonlinear mutual information across local, mesoscopic and global topological scales. First, we have motivated the comparison of the two types of networks at equal edge densities. We have considered only low edge densities, that were shown to yield networks containing statistically highly significant edges as established on the basis of various significance tests. It has been then consistently shown for AOGCM and reanalysis surface air temperature data, that the networks agree well on the local and mesoscopic topological scales. Using the surface pressure field to construct climate networks also yielded qualitatively similar results and identical conclusions on these scales. For the surface air temperature data sets, we have found some interesting qualitative and quantitative deviations at the global scale using betweenness centrality. Even though there still is a high degree of similarity, the deviations are highly localized and structured pointing at a possible involvement of nonlinear processes in their formation.

This work also demonstrates, that our method of calculating mutual information for relatively short time series is reliable at least for the strongly linear interrelations detected by the Pearson correlation coefficient. The global topological scale is of particular interest, since it opens novel perspectives for the understanding of climatological phenomena. For example, as applied to the climate networks discussed in this article, betweenness centrality allows to measure the importance of localized regions on the earth's surface for the transport of dynamical information within a climatological field in the long term mean \cite{Donges2008}. Further work is needed to establish, whether the observed deviations on the global topological scale could be due to nonlinear physical processes in the climate system, that are only detectable using mutual information. In the future, we plan to assess this problem by constructing climate networks using a novel method based on statistical significance, i.e. by adding edges to the climate network depending on the significance level of the correlation measure with respect to reasonable null hypotheses. One could then identify candidates for nonlinear interrelationships as edges that have an associated significant mutual information and a Pearson correlation that is not significant.


\begin{acknowledgement} 
We acknowledge the modeling groups, the Program for Climate Model Diagnosis and Intercomparison (PCMDI) and the WCRP's Working Group on Coupled Modelling (WGCM) for their roles in making available the WCRP CMIP3 multi-model data set. Support of this data set is provided by the Office of Science, U.S. Department of Energy. We also acknowledge the use of NCEP Reanalysis Derived data provided by the NOAA/OAR/ESRL PSD, Boulder, Colorado, USA, from their Web site at http://www.cdc.noaa.gov/. We thank Andr\'{e} Bergner, Gorka Zamora-L\'{o}pez, Anders Levermann and Gunnar Schmidt for stimulating discussions and helpful comments. Furthermore, we express our gratitude to Reik Donner and two anonymous referees for their valuable suggestions. J\"urgen Kurths acknowledges the SFB 555 (DFG) for financial support.
\end{acknowledgement}


\end{document}